\documentclass[sn-mathphys-num]{sn-jnl}

\usepackage{graphicx}%
\usepackage{multirow}%
\usepackage{amsmath,amssymb,amsfonts}%
\usepackage{amsthm}%
\usepackage{mathrsfs}%
\usepackage[title]{appendix}%
\usepackage{xcolor}%
\usepackage{textcomp}%
\usepackage{manyfoot}%
\usepackage{booktabs}%
\usepackage{algorithm}%
\usepackage{algorithmicx}%
\usepackage{algpseudocode}%
\usepackage{listings}%
\usepackage{subcaption}
\usepackage{overpic}
\usepackage[version=4]{mhchem}

\usepackage[table]{xcolor}

\definecolor{accdarkgreen}{RGB}{90,190,115}
\definecolor{accgreen}{RGB}{144,238,144}
\definecolor{accgreenyellow}{RGB}{200,230,120}
\definecolor{accyellow}{RGB}{255,235,100}
\definecolor{accorange}{RGB}{255,180,90}
\definecolor{accred}{RGB}{255,120,120}

\usepackage[english]{babel}

\usepackage{amsmath}
\usepackage{graphicx}

\usepackage{algorithm}
\usepackage{algorithmicx}
\usepackage{algpseudocode}
\usepackage{multicol}
\usepackage{amsmath}
\usepackage{fancyvrb} 
\usepackage{tcolorbox} 

\usepackage{algpseudocode}
\usepackage{todonotes}
\usepackage[acronym]{glossaries}
\newacronym{amr}{AMR}{adaptive mesh refinement}
\newacronym{cfd}{CFD}{computational fluid dynamics}
\newacronym{dsmc}{DSMC}{direct simulation Monte carlo}
\newacronym{gpu}{GPU}{graphics processing unit}
\newacronym{cpu}{CPU}{central processing unit}
\newacronym{hpc}{HPC}{high performance computing}
\newacronym{pde}{PDE}{partial differential equation}
\newacronym{nlr}{NLR}{National Laboratory of the Rockies}
\newacronym{doe}{DOE}{Department of Energy}
\newacronym[longplural={Gaussian processes}]{gp}{GP}{Gaussian Process}
\newacronym{ac}{AC}{Adaptive Computing}
\newacronym{smt}{SMT}{Surrogate Modeling Toolbox}
\newacronym{uq}{UQ}{uncertainty quantification}
\newacronym{amar}{AMAR}{Adaptive Mesh and Algorithm Refinement}
\newacronym{kmc}{KMC}{kinetic Monte Carlo}
\newacronym{spparks}{SPPARKS}{Stochastic Parallel PARticle Kinetic Simulator}
\newacronym{dft}{DFT}{density functional theory}
\newacronym{mpi}{MPI}{Message Passing Interface}
\newacronym{qoi}{QOI}{quantity of interest}
\newacronym{rom}{ROM}{reduced-order model}
\newacronym{mf}{MF}{multi-fidelity}
\newacronym{vasp}{VASP}{Vienna Abi Initio Simulation Package}

\newcommand{\HeatFlux}{\boldsymbol{\mathcal{Q}}}
\newcommand{\SpeciesFlux}{\boldsymbol{\mathcal{F}}}

\newcommand{\StressTensor}{\boldsymbol{\Pi}}

\makeglossaries
\glsdisablehyper
\glsenableentrycount

\theoremstyle{thmstyleone}%
\theoremstyle{thmstyletwo}%

\theoremstyle{thmstylethree}%

\begin{document}

\title{Autonomous atomistic–continuum coupling through uncertainty-aware active learning}

\author[1]{\fnm{Daniel} \sur{Abdulah}}\email{dabdulah@mit.edu}
\author[2]{\fnm{Hariswaran} \sur{Sitaraman}}\email{Hariswaran.Sitaraman@nlr.gov}
\author[2,3]{\fnm{Ross E.} \sur{Larsen}}\email{Ross.Larsen@nlr.gov}
\author*[2]{\fnm{Kevin Patrick} \sur{Griffin}}\email{Kevin.Griffin@nlr.gov}
\author[4]{\fnm{Grace} \sur{Wei}}\email{gawei@berkeley.edu}
\author[5,6]{\fnm{Janelle} \sur{Domantay}}\email{Janelle\_Domantay@mines.edu}
\author[2]{\fnm{Ethan} \sur{Young}}\email{Ethan.Young@nlr.gov}
\author[2]{\fnm{Marc T.} \sur{Henry de Frahan}}\email{marchdf@gmail.com}

\affil[1]{\orgdiv{Department of Earth, Atmospheric, and Planetary Sciences}, \orgname{Massachusetts Institute of Technology}, \orgaddress{\street{77 Massachusetts Avenue}, \city{Cambridge}, \postcode{02139}, \state{MA}, \country{USA}}}
\affil[2]{\orgdiv{Computational Science Center}, \orgname{National Laboratory of the Rockies}, \orgaddress{\street{15013 Denver West Parkway}, \city{Golden}, \postcode{80401}, \state{CO}, \country{USA}}}
\affil[3]{\orgdiv{Renewable and Sustainable Energy Institute}, \orgname{University of Colorado Boulder}, \orgaddress{\street{2055 Regent Dr}, \city{Boulder}, \postcode{80309}, \state{CO}, \country{USA}}}
\affil[4]{\orgdiv{Department of Materials Science and Engineering}, \orgname{University of California Berkeley}, \orgaddress{\street{210 Hearst Memorial Mining Building}, \city{Berkeley}, \postcode{94720}, \state{CA}, \country{USA}}}
\affil[5]{\orgdiv{Department of Computer Science}, \orgname{University of Illinois Urbana-Champaign}, \orgaddress{\street{801 South Wright Street}, \city{Urbana}, \postcode{61801}, \state{IL}, \country{USA}}}
\affil[6]{\orgdiv{Department of Mechanical Engineering}, \orgname{Colorado School of Mines}, \orgaddress{\street{1500 Illinois St}, \city{Golden}, \postcode{80401}, \state{CO}, \country{USA}}}


\abstract{We develop an uncertainty‑aware surrogate modeling approach for coupling an atomistic kinetic Monte Carlo (KMC) surface‑chemistry solver with a continuum multi‑species fluid dynamics solver across disparate physical length and time scales. The surrogate model predicts boundary mass fluxes for six species and their uncertainties, which the continuum solver monitors to automatically trigger additional KMC simulations when new training data are needed. This enables efficient multiscale coupling and maintains a user‑specified level of accuracy. We demonstrate the approach for reactive halide‑perovskite vapor deposition, which involves a multi‑species vapor jet impinging on a reactive substrate. The KMC solver uses a DFT‑calibrated energetics model that we construct as part of this work and provides net deposition and desorption rates for six species. With a conservative uncertainty threshold, the surrogate-coupled approach reproduces cumulative mass fluxes of the directly coupled approach with errors typically within 10–20\%, while capturing the non‑monotonic dependence of net deposition on species partial pressure. Importantly, the results show that detailed crystal structure need not be explicitly represented to predict quantities exchanged between atomistic and continuum solvers; species‑resolved mass fluxes provide a sufficiently low‑dimensional representation of the atomistic response. For this accuracy target, the autonomous online‑learning approach reduces the number of required atomistic simulations by two to three orders of magnitude relative to direct multiscale coupling.}




\maketitle

\section{Introduction} \label{sec:introduction}

Fundamental and applied science investigations pertaining to many engineering technologies are often ``multiscale'' in nature. There is a need to resolve a large range of scales from the atomistic level, on the order of nanometers (molecules), to the continuum level, on the order of meters (devices and reactors), to accurately understand many connected physical phenomena. The resolution of phenomena at various length scales is typically done in isolation. The coupling between scales, when it happens, is often through user-defined reduced-order models that propagate information up from smaller to larger length scales. Such coupling is often one-sided, i.e., it parametrizes (models) molecular phenomena for continuum-level solvers, and offline, i.e., coupling across scales is not performed at every simulation time-step but rather is performed once as two separate workflows. Such models are only valid over the subset of the input space where training data was collected. However, this subspace is often different than the subspace where the model will be queried by the continuum solver since the latter is typically not known \emph{a priori}. As a result, the continuum solver often uses the model in an extrapolative manner, which introduces epistemic uncertainty. 

Atomistic simulations, which resolve phenomena at atomic or molecular length scales, typically involve spatial scales on the order of nanometers and, for methods such as molecular dynamics, time scales on the order of femtoseconds to picoseconds. Meanwhile, continuum simulations are typically associated with significantly longer length and time scales that depend on the size of the device under consideration (e.g., $\sim$ ms in a 1 m size reactor). The separation of time and length scales can make direct coupling of these simulations require an intractable number of timesteps and spatial resolution to capture the necessary range of scales.

The primary case study for this work is the problem of vapor deposited film growth for halide perovskites \cite{10.1039/c5ta08963h,hoerantner2019high,bonomi2020physical,raiford2020applications,guesnay2021vapor}. Typical simulations of vapor-driven growth may be atomistic in nature and consider crystal growth at atomistic scales \cite{kalke2001kinetic,battaile2002kinetic,yang2009ge,shirazi_atomistic_2014,elliott_modeling_2016} or consider reactor-scale phenomena typically using continuum-scale simulations relying on parametric or phenomenological models for surface growth \cite{yang2017numerical}. 
The \gls{kmc} approach \cite{kalke2001kinetic,battaile2002kinetic,shirazi_atomistic_2014,elliott_modeling_2016} already represents a coarse-graining approximation from molecular dynamics \cite{yang2009ge} to rare events, yielding a $\sim$ 1000-fold increase or more in the typical timestep. Nonetheless, direct coupling of atomistic to continuum simulations remains a challenge for large times because the atomistic, e.g., \gls{kmc}, part of the simulation can be thousands of times slower than the continuum steps, making times associated with large length scales impossible to achieve.

One application where accounting for atomistic scale physics does not require resorting to multi-solver coupling is the field of fluctuating hydrodynamics. In this context, the Navier-Stokes equations are insufficient because atomistic stochasticity significantly affects the continuum solution, as in Brownian motion or in moderately rarefied micro- and nanoscale flows. The Landau–Lifshitz Navier-Stokes (LLNS) equations incorporate atomistic physics in continuum simulations via a thermal noise term, which is derived analytically from statistical mechanics (e.g., \cite{donev2010accuracy,donev2014low,bell2010computational,balboa2012staggered}). As a result, LLNS can capture fluctuation-driven phenomena without requiring direct computational coupling with atomistic simulations.

In a study by Harris \citep{harris2025validation}, adaptive mesh refinement is used alongside a cell-by-cell decision to use a continuum or rarefied discrete velocity method based on a rarefied gas criterion. This is an example where two solvers are coupled in situ rather than through a surrogate model.

In most multiscale applications, analytical treatments of the microphysics are not available, so numerical microscale solvers are introduced and must be coupled to the continuum solver. The Multiscale Universal Interface is a software library for directly coupling atomistic and continuum solvers \cite{tang2015multiscale}. 
The National Aeronautics and Space Administration (NASA) Multiscale Analysis Tool is an example of a modular software framework for multiscale modeling and has been deployed with direct and offline surrogate coupling for a variety of materials simulation applications \cite{arnold2023enabling}.

\begin{figure}[!htbp]
    \centering
    \includegraphics[width=0.8\linewidth]{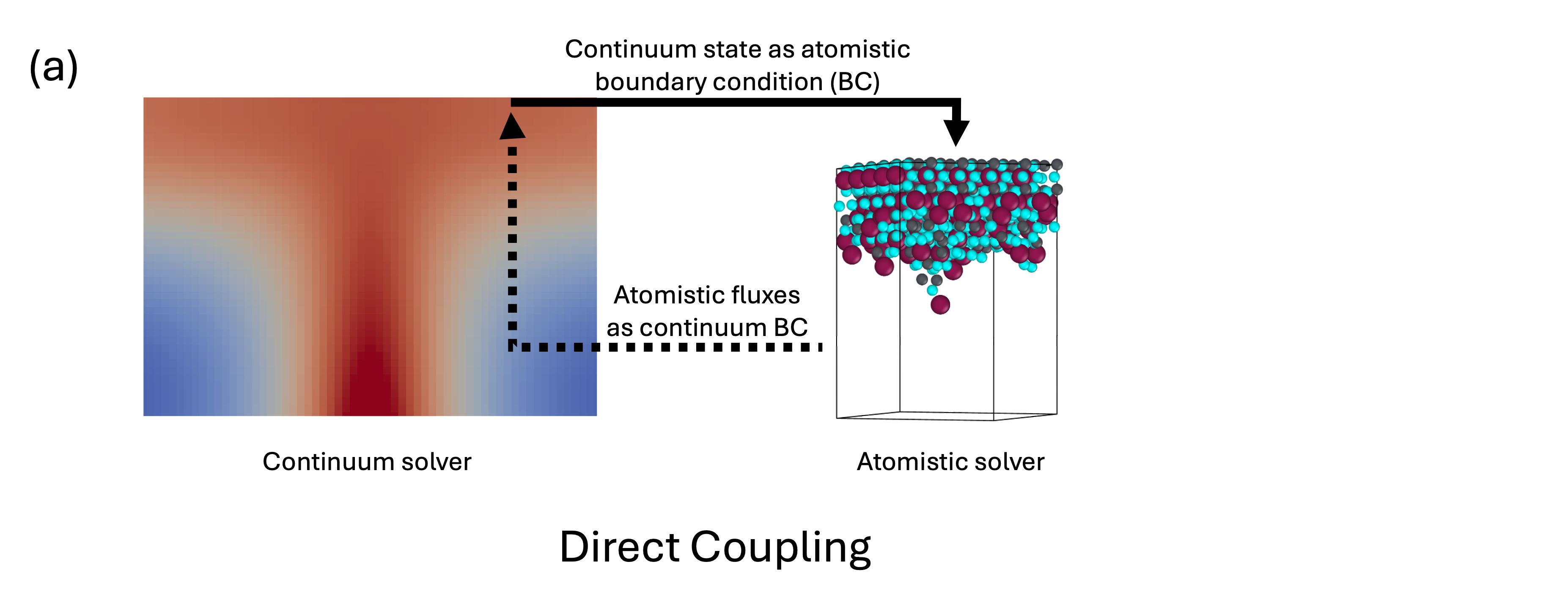}
    \includegraphics[width=0.8\linewidth]{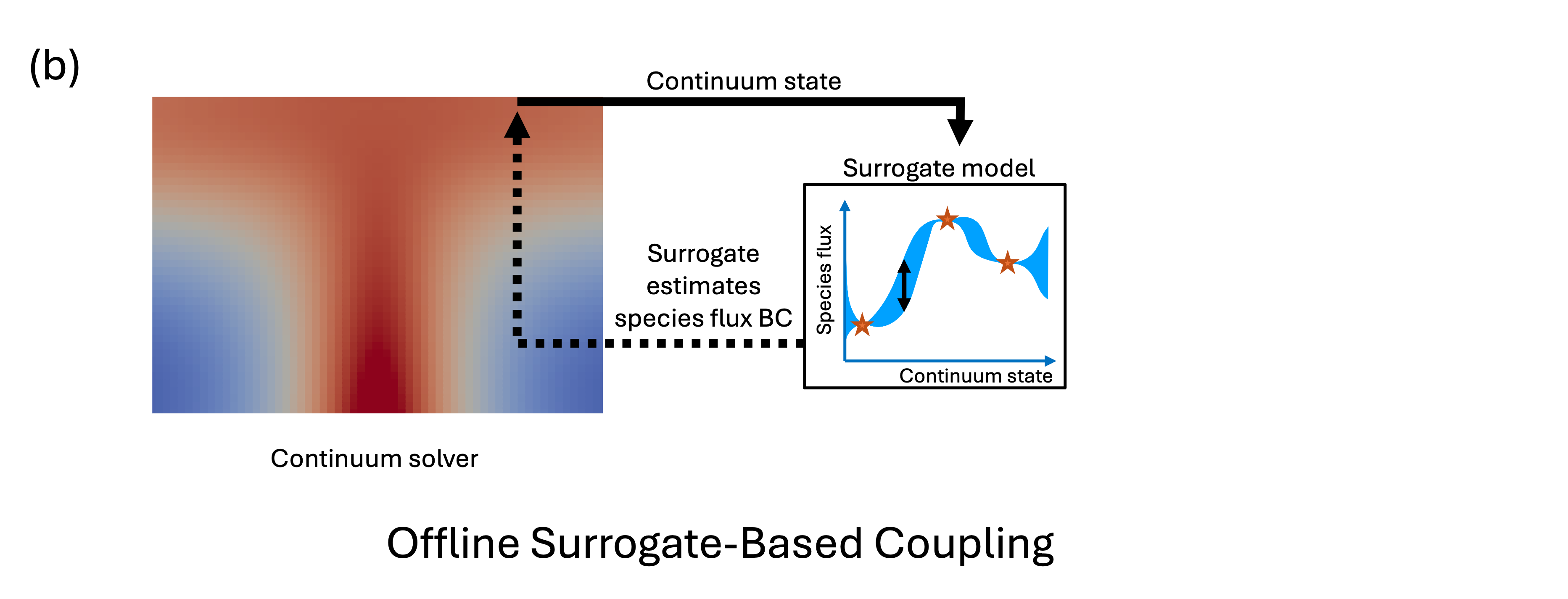}
    \includegraphics[width=0.8\linewidth]{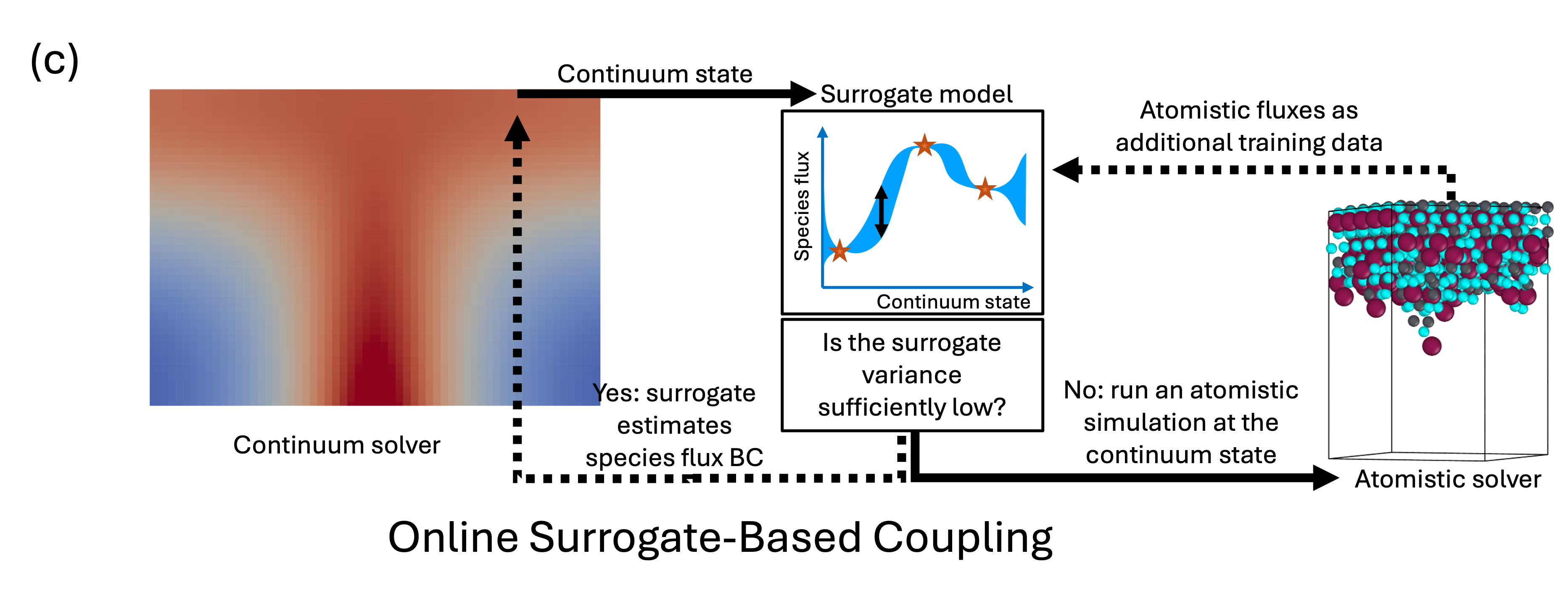}
    \caption{Three strategies for coupling continuum and atomistic solvers. (a) \textbf{Direct coupling-} the continuum solver queries the atomistic solver at predefined discrete instances in time and space. (b) \textbf{Offline surrogate-based-} a surrogate model is trained {\it a priori} using atomistic calculations, then the static surrogate model is used as a microphysics closure for the continuum solver. (c) \textbf{Online surrogate-based-} (the main focus of this work) the surrogate model is retrained/improved with data from additional microphysics simulations as the continuum solver proceeds.}
    \label{fig:summary}
\end{figure}

Surrogate (statistical) modeling presents an opportunity to couple atomistic- and continuum-scale simulations at an affordable computational cost compared to direct coupling. In Fig.~\ref{fig:summary}, we consider three paradigms for coupling these solvers for our perovskite synthesis application where an atomistic solver describes detailed boundary physics. Fig.~\ref{fig:summary}(a) depicts direct coupling, for which an atomistic simulation is performed for each point on the boundary and at each timestep of the continuum solver, using the local continuum state as a boundary condition for the atomistic solver and using an average of the atomistic solver's solution to provide a boundary condition for the continuum solver. This strategy often involves significant wasteful recomputation as the atomistic solver queries are likely correlated in space and time due to corresponding correlations in the continuum state. Fig.~\ref{fig:summary}(b-c) depict surrogate-based coupling, for which a surrogate model is trained on data from atomistic simulations. At each point on the boundary and at each timestep of the continuum solver, a query for the boundary flux is made to a surrogate model instead of to the atomistic solver. In the offline approach of Fig.~\ref{fig:summary}(b), the surrogate model is trained by running the atomistic simulations at prescribed operating conditions before the continuum solver begins executing. The training data may not correspond to the states actually visited by the continuum solver, requiring the surrogate model to extrapolate beyond its training data. In the online surrogate-based approach shown in Fig.~\ref{fig:summary}(c), each instance of querying the surrogate model implies a decision point. If the variance (uncertainty) of the surrogate is sufficiently low, then the surrogate's estimates of the species fluxes are used as in the offline approach. If not, an atomistic simulation is conducted at the relevant continuum state, and the solution is used to update the surrogate model. In this online learning approach, the surrogate can be arbitrarily accurate by reducing the variance tolerance, and can recover the behavior of the direct-coupling and offline-only approaches in the limit of zero variance tolerance and infinite variance tolerance, respectively, along with the incurred computational cost.
Compared to the direct coupling approach, online learning is potentially more computationally efficient since rather than calling the atomistic simulation at highly correlated operating conditions, data from previous queries is interpolated.

In some applications, it is a challenge to parameterize and sample the surrogate model adequately. That is, the microscale solver has a prohibitively large number of degrees of freedom, and attempting to summarize them in a low dimensional way such that surrogate model can be learned and deployed with the desired level of confidence is challenging. If the surrogate model has too many dimensions, it will be expensive to run enough micro-simulations to collect enough training data to achieve statistical convergence of the surrogate. Conversely, if the surrogate model is under-parameterized, its accuracy will degrade. 
An example of striking this balance is given by Tatsios et al. \cite{tatsios2025dsmc}, where a continuum \gls{cfd} simulation code is coupled to a micro scale \gls{dsmc} code. The \gls{dsmc} code is tasked with estimating coefficients in the constitutive equation and boundary conditions. In this way, an assumption was made about the constitutive and boundary models to use for rarefied flows solved with \gls{cfd}. This allows for the high dimensional \gls{dsmc} solution to be modeled with a small number of model coefficients. Similarly, in the present work, the \gls{kmc} contains high dimensional data describing the crystal occupancy, but only the mass fluxes are predicted by the surrogate. 

Next, we mention various other examples of multisolver coupling in the literature and classify them as direct, offline, or online coupling. Xiao and Ni \citep{xiao2020multiscale} perform offline surrogate-based coupling of \gls{kmc} surface etching simulations to a continuous plasma solver using a neural network as the surrogate model. 
The work of Rocha, Kerfriden, and Der Meer \cite{rocha2020fly} involves online surrogate-based coupling of a continuum-scale finite element solver to a microstructural finite element solver through the constitutive equation for the continuum solver. 
In the work of Leiter et al. \cite{leiter2018accelerated}, a continuum finite element solver is coupled through an online surrogate model to an atomistic scale dissipative particle dynamics model through the equation of state. 
Stephenson, Kermode, and Lockerby \cite{stephenson2018accelerating} employs an online surrogate model to communicate between discrete and continuum scales of molecular fluid flow.

Existing studies have demonstrated online surrogate models that use predictive uncertainty to adaptively trigger new microscale simulations for continuum–microscale coupling. However, these approaches have not been applied to atomistic surface chemistry coupled to a multi‑species reactive flow through species‑resolved boundary mass fluxes. In this work, we develop a fully autonomous, uncertainty‑aware boundary‑flux surrogate that couples a continuum multi‑species CFD solver to an atomistic KMC surface‑chemistry solver. We further construct an energetics model for the KMC solver—calibrated using DFT simulations—to enable accurate treatment of halide‑perovskite vapor deposition. The surrogate predicts species‑resolved mass fluxes along with their variances, and the CFD solver monitors these variances to trigger additional KMC simulations only when necessary, retraining the surrogate online. Applied to reactive halide‑perovskite vapor deposition involving six species, the approach reproduces the expected non‑monotonic flux–pressure behavior while reducing the number of atomistic simulations by two to three orders of magnitude with 10-20\% error. Importantly, the results demonstrate that the detailed crystal structure does not need to be explicitly represented by the surrogate to accurately predict the quantities exchanged across scales; instead, species‑resolved mass fluxes provide a sufficiently low‑dimensional representation of the atomistic response, substantially reducing the computational cost of multiscale coupling.

The manuscript is organized as follows: Section \ref{sec:kmc} describes the atomistic \gls{kmc} solver and energetics model. Section \ref{sec:coupling} describes the continuum \gls{cfd} solver and its direct coupling to the atomistic model. Section \ref{sec:surrogate} presents the surrogate model and indirect coupling through online learning. Results and conclusions are discussed in Sections \ref{sec:results} and \ref{sec:conclusions}, respectively.

\section{Atomistic modeling of vapor deposition} \label{sec:kmc}

In this section, we present the atomistic model for the vapor deposited crystal growth of a perovskite crystal having the generic stoichiometry $ABX_3$. All-inorganic halide perovskite crystals such as $CsPbI_3$ have been of recent interest for photovoltaic and optoelectronic applications \cite{sutton2018cubic,zhao2018thermodynamically,wang2019thermodynamically,li2025density}. To simulate vapor deposition, we implemented the \gls{kmc} algorithm described in detail below as a new, independent application in the \texttt{lattice} class of the \gls{spparks} \gls{kmc} simulation code \cite{slepoy_constant-time_2008,webb_edmund_blackburn_iii_crossing_2009}. The algorithm incorporates key deposition, growth, and diffusion processes relevant for vapor deposition as described below. It follows the general approach for atomic layer deposition simulations described by Elliott and coworkers \cite{shirazi_atomistic_2014,elliott_modeling_2016}. The crystal growth algorithm we implemented can be applied to vapor deposition of arbitrary crystal structures given user definitions of lattice type, identification of site neighbor lists, species types, event rates, and energies associated with a site being occupied. Below we will generically refer to the species in a perovskite crystal structure as $A$, $B$ and $X$, with crystal stoichiometry, $ABX_3$.

\subsection{Lattice growth model}

The lattice model assumes that atoms and molecules in the gas-phase continuum can deposit on a surface on the sites of a pre-defined three-dimensional lattice, making this an ``on-lattice'' \gls{kmc} method \cite{webb_edmund_blackburn_iii_crossing_2009}. With the assumption of growth on a pre-defined lattice, our method cannot determine what structure might grow under a given set of deposition conditions, only how growth that is restricted to the occupancy of empty sites in a prescribed structure might occur, under given conditions. Off-lattice methods \cite{henkelman2001long} do not restrict atom locations but the extra degrees of freedom associated with defining structures without an underlying lattice requires a higher spatial fidelity for the potential energy, leading to a smaller speedup going from molecular dynamics to \gls{kmc}. Our purpose in this paper is to demonstrate a multi-scale capability via online surrogate modeling, so we proceed with a specified orthorhombic lattice and (001) surface that is known to be stable for the $CsPbI_3$ system \cite{sutton2018cubic,zhao2018thermodynamically}. The fixed choice of lattice also includes fixing the crystal parameters to match the experimental bulk parameters,\cite{sutton2018cubic} so lattice strain for a partially-grown crystal is neglected in this simplified model.

The growth model allows the following processes: deposition from the gas onto the surface of the growing crystal; desorption of ions or molecules from the surface; diffusion of ions in the growing crystal, both on the surface and in and out of the bulk. The rates for these events are computed from the kinetic theory of gases (for deposition) or from transition state theory for all other processes. The model is specified by a list of possible species in the gas phase that can impact the surface, at rates given by the kinetic theory of gases as described in Appendix \ref{app:KMC}. These can be neutral atoms or molecules, or ions or molecular fragments. When a molecule deposits on the surface, we cease thinking of it as a molecular unit and describe it as a number of independent ions occupying distinct lattice sites. For example during the synthesis of $ABX_3$ (e.g., $CsPbI_3$), when an AX molecule is deposited from the gas phase onto an unoccupied lattice site, that lattice site is switched to being occupied by an $A^+$ or $X^-$ ion and simultaneously a neighboring unoccupied site becomes occupied by an $X^-$ or $A^+$ ion, respectively. Desorption events can involve a single ion or molecule leaving the surface, which is enacted by changing the occupancy of the relevant lattice sites to \texttt{unoccupied}. A \emph{molecular} desorption event involves converting a pair (AX or BX) or triple (BX$_2$) of neighboring occupied sites to the unoccupied state. We avoid double counting events when we form the list of possible desorption events by associating the molecular desorption events with the $A^+$or $B^+$ occupied site. Our model also allows multiple ways for a molecule to desorb based on desorption of a given A or B atom, provided there are sufficient X-occupied neighbors that a molecule could form in more than one way. For example, if some A ion had two X ions as neighbors, denoted here X$_1$ and X$_2$, there would be separate rates for AX$_1$ and AX$_2$ to desorb. In all of our \gls{kmc} runs, we assume that deposition and desorption events occur rarely enough that the temperature of the substrate remains constant.

The full list of atomistic processes that are included in the atomistic model for growth of an $ABX_3$ perovskite crystal from vapor deposition is listed in Appendix \ref{app:KMC}, along with a brief description of the KMC algorithm used for time-stepping. Energies associated with adsorption and desorption events are given by a model for ion energies as a function of the local neighborhood that was fitted to adsorbate and defect energies computed from electronic structure calculations based on density functional theory, as described in Appendix \ref{app:energy_model}.

\subsection{SPPARKS software}
The \gls{kmc} model presented in this paper is implemented as an application in the Stochastic Parallel PARticle Kinetic Simulator, SPPARKS, software~\cite{webb_edmund_blackburn_iii_crossing_2009}. SPPARKS is an open-source software package that is used for on-lattice and off-lattice models for \gls{kmc}. It is readily extensible with user additions of applications, as is done for this work. SPPARKS is developed using the C++ programming language and, as such, it is straightforward to link to the SPPARKS library to obtain and interact with an object representing the \gls{kmc} model in other applications written using C or C++. From a software engineering standpoint, the C++ framework presented in Section \ref{sec:coupling} leverages this code coupling strategy to create the \gls{kmc} object in memory and interact with it through member function calls during the simulation run time. 

\section{Multi-scale direct coupling to continuum solver} \label{sec:coupling}

In this section, we present the continuum model and the multiscale modeling strategy for coupling the continuum solver to the atomistic model presented in Section~\ref{sec:kmc}. This section covers the direct coupling paradigm depicted in Fig.~\ref{fig:summary}(a).

\subsection{Continuum reacting flow model} \label{sec:continuum}
The compressible reacting-flow solver in this work solves the conservation equations for mass, species mass fractions, momentum, and energy as shown below:
\begin{align}
&\frac{\partial }{\partial t} \left( \rho \right)  + { \nabla} \cdot \left( \rho \mathbf{u} \right) = 0,\label{eqn:cont}\\
&\frac{\partial }{\partial t} \left( \rho Y_k \right)  + { \nabla} \cdot \left( \rho \mathbf{u} Y_k \right) = -{ \nabla} \cdot \SpeciesFlux_k \label{eqn:spec}\\
&\frac{\partial }{\partial t} \left( \rho \mathbf{u} \right)  + { \nabla} \cdot \left( \rho \mathbf{u} \otimes \mathbf{u}  \right) + { \nabla} p = {\nabla} \cdot   \StressTensor, \label{eqn:mom}\\
& \frac{\partial }{\partial t} \left( \rho E \right)  + { \nabla} \cdot \left( \rho \mathbf{u} E + p \mathbf{u}  \right) = { \nabla} \cdot  { \HeatFlux}. \label{eqn:energy}
\end{align}
where $\rho$ is the density, $\mathbf{u}$ is the velocity vector, $p$ is the pressure of the mixture, and $Y_k$ is the mass fraction of the $k$-th species. Thermal equilibrium is assumed for the different species in the gaseous mixture, at a common temperature, $T$. The boundary conditions to the above equations are outlined in Figure~\ref{fig:boundary_conditions} (discussed in Section~\ref{sec:results}) and apply to all dynamical fields. The exact treatment of the substrate boundary is described in Section \ref{sec:coupling-algorithm}.

The viscous stress tensor, $\StressTensor$, under the Newtonian assumption, is given by:
\begin{equation}
\StressTensor = \eta \left( \nabla \mathbf{u} + ( \nabla \mathbf{u} )^T\right) + \left( \kappa - \frac{2}{3} \eta\right) \mathbf{I} \nabla \cdot \mathbf{u}
\end{equation}
where $\eta$ is the shear viscosity and $\kappa$ is the bulk viscosity. $\SpeciesFlux_k=-D_k \rho \nabla Y_k$ is the diffusive transport flux of the $k$-th species with diffusion coefficient $D_k$, approximated using a mixture averaged diffusion process, and $\HeatFlux$ is the thermal conduction heat flux. 
The system of equations is closed using the ideal gas equation-of-state. Gas phase chemistry, though possibly of interest in the application considered here, is not included for simplicity.

Molecular and chemical parameters used for the six species that are typical of pervoskites of the form $\mathrm{ABX_3}$ are provided in Table \ref{tab:species}. We use parameters appropriate for the deposition of $\mathrm{CsPbI_3}$ in this work with the dominant background gas being $\mathrm{N_2}$. The viscosity of $\mathrm{N_2}$ used is 1.788e-5~Pa~s \cite{johnston1951viscosities}.

\begin{table}[]
    \centering
    \begin{tabular}{l|c|c|c|c}
        species & mol wt.\ (g/mol) & $\gamma$ & atomic/molecular size (\r{A}) & $D^0_k\,\mathrm{(m^2/s)}$ \\
         \hline
        A ($\mathrm{Cs^+}$) & 133 & 1.67 & 3.49 & 1.70e-5  \\
        B  ($\mathrm{Pb^{2+}}$) & 206 & 1.67 & 3.00 & 1.65e-5 \\
        X  ($\mathrm{I^-}$) & 127 & 1.67 & 5.33 & 7.34e-6 \\
        AX ($\mathrm{CsI}$) &  260 & 1.40 & 5.50 & 7.00e-6 \\
        BX ($\mathrm{PbI^+}$)&  333 & 1.40 & 5.30 & 6.60e-6  \\
        BX$_2$ ($\mathrm{PbI_2}$) & 460 & 1.33 & 6.30 & 4.90e-6 \\
        \ce{N2} & 28.0 & 1.40 & 1.10 & 1.89e-5
    \end{tabular}
    \caption{Properties of the six gas species assuming the dominant background gas to be \ce{N2}. The diffusion coefficient for \ce{N2} here is its self-diffusion coefficient.}
    \label{tab:species}
\end{table}
The diffusion coefficient for each of the species ($D^0_k$) at atmospheric pressure in Table \ref{tab:species} are obtained by calculating the mutual diffusion coefficients with the background gas, $\mathrm{N_2}$, using the modified  Chapman-Enskog formula from Yang et al.\ \cite{yang2017numerical}. The atomic/molecular size (Table \ref{tab:species}) and the self diffusion coefficient of $\mathrm{N_2}$ (1.89e-5 $\mathrm{m^2/s}$ \cite{burgess2024self}) are used in the calculation of the mutual diffusion coefficients. A simple pressure scaling is applied to get diffusion coefficient at typical low pressure processing conditions as $D_k=D_0 (p_{atm}/p)$.

The spatial discretization follows a second-order discretization for hyperbolic fluxes using the AUSM+up~\cite{liou2006sequel} scheme and a central difference scheme for the diffusion terms. The transport equations are solved using a global second-order Runge-Kutta time-stepping framework. Throughout this work, the timestep is set such that the maximum Courant number $\le 0.1$. Central to the solver is the use of the block-structured Cartesian \gls{amr} library, AMReX \cite{zhang2019amrex}, which allows the solver to locally achieve the resolution necessary to simulate complex physical phenomena, such as at regions of high vorticity or at pervoskite growth interfaces. This library has been used in a wide range of applications and demonstrated performance efficiency on high performance computing architectures~\cite{henrydefrahan2022, PeleLMeX_JOSS, HenrydeFrahan2024, Sitaraman2021, sitaraman2021b, prfwimer2023, malaya2023,Sharma2023,amrwind2025}.

\subsection{Multi-scale coupling algorithm} \label{sec:coupling-algorithm}

The problem of perovskite lattice formation with deposition from an impinging jet requires coupling our continuum model of fluid dynamics (Section \ref{sec:continuum}) and our atomistic model (Section \ref{sec:kmc}) at the perovskite growth boundary layer as indicated in Fig.~\ref{fig:summary}(a). Our approach is summarized in Algorithm~\ref{alg:directly_coupled}. At the atomistic scale, the movement of ions on and off of the perovskite crystal depends on continuum state, namely the partial pressures of each ion species $pY_k$ and the local temperature $T$. Conversely, the continuum mass fractions of each species (Equations \ref{eqn:spec}) are informed by gas flux into and out of the crystal. The continuum timestep thus involves an intermediate step in which each grid cell bordering the substrate passes its state to the atomistic model to obtain species fluxes as a boundary condition. 
The atomistic model employs \gls{kmc} simulations (Section \ref{sec:kmc}). These consider a 3D subdomain consisting of a $N_x\times N_y$ doubly periodic horizontal patch on the substrate, along with $N_z$ sites for a crystal to grow vertically. The \gls{kmc} subdomain is considered to be a small fraction of one continuum grid cell along the substrate to ensure tractable computational costs. The net number of atoms deposited in the pervoskite crystal is calculated in the microgrid subdomain, and from this, a species flux for the continuum simulation is obtained as shown in Algorithm~\ref{alg:directly_coupled}. 

\begin{algorithm}
\caption{Directly coupled continuum–atomistic simulation for perovskite growth}
\label{alg:directly_coupled}
\begin{algorithmic}[1]
\State \textbf{Inputs:} 
\State \hspace{1em} Ensemble size $N_{\text{ens}}$
\State \hspace{1em} KMC timestep $\Delta t_m$
\State \hspace{1em} Transient time to discard $t_{\text{trans}}$
\State \hspace{1em} Fluid timestep $\Delta t$
\State \hspace{1em} Continuum/atomistic area ratio $A_M / A_m$
\State \hspace{1em} Avogadro’s number $N_A$, molar masses $M_k$ for all species $k$
\For{each continuum timestep}
    \For{each grid cell adjacent to the substrate}
        \State Extract local state: partial pressures $\{pY_k\}$ and temperature $T$ from continuum
        \For{$i = 1$ to $N_{\text{ens}}$}
            \State Initialize KMC domain with $T$, $\{pY_k\}$
            \State Run KMC simulation for time $t_\mathrm{trans}+\Delta t_m$
            \State Discard first $t_{\text{trans}}$ of data (transient phase)
            \State Record net deposited atoms $\text{net}_k^{(i)}$ for each species $k$
        \EndFor
        \State Compute ensemble-averaged deposition: 
        \[
        \overline{\text{net}_k} = \frac{1}{N_{\text{ens}}} \sum_{i=1}^{N_{\text{ens}}} \text{net}_k^{(i)}
        \]
        \State Compute species flux to continuum model using:
        \[
        \frac{\partial Y_k}{\partial t} = 
        \frac{\overline{\text{net}_k} \cdot A_M}
        {A_m \cdot N_A \cdot M_k \cdot \Delta t_m}
        \]
        \State Apply species flux as boundary condition in continuum model
    \EndFor
    \State Advance continuum model by one timestep $\Delta t$
\EndFor
\end{algorithmic}
\end{algorithm}

When an atomistic simulation is called, we aim to reduce stochasticity in the output of the \gls{kmc} results. As observed in Section~\ref{sec:results}, the net rate of deposition and desorption of species onto a clean substrate is initially governed by random rare events before becoming nearly constant in time as an asymptotically rough wetted crystal surface develops above the substrate. When conducting the \gls{kmc} simulations, we disregard this initial stage $t_\mathrm{trans}$ and only retain outputs from the approximately steady state period of duration $\Delta t_{m}$. We conduct an ensemble of $N_\mathrm{ens}$ KMC simulations and average the resulting deposition rates, which is equivalent to sampling multiple KMC patches within the continuum grid cell. The averaged $\text{net}_k$ is passed back to the continuum solver as a boundary condition as given by the definition of $\text{net}_k$ and the equation for $\partial Y_k /\partial t$ in Algorithm~\ref{alg:directly_coupled}.

\section{Surrogate modeling} \label{sec:surrogate}
The cost of the directly coupled simulations is dominated by the cost of the atomistic KMC calculations. When KMC calculations are turned off (i.e. for a simulation with surface reactions disabled), the simulation runs 3500x faster. 

We expect that the average of the atomistic simulation trials conducted in directly coupled simulations is highly correlated because the input conditions (species concentrations and temperature) for the atomistic simulations are taken from the continuum simulation solution, which is continuous in time and space. This motivates the use of a surrogate-based multiscale coupling to interpolate between prior correlated atomistic simulations to avoid unnecessary computation. The approach is outlined visually in Fig.~\ref{fig:summary}(c) and in pseudo code in Fig.~\ref{fig:algo_comparison}(b).

\begin{figure}[ht]
\begin{tcolorbox}[colback=white, boxrule=0.3pt]
\begin{minipage}[t]{0.48\textwidth}
\begin{algorithm}[H]
\caption*{\textbf{(a) Directly coupled (simplified)}}
\begin{algorithmic}[1]
\State \textbf{Input:} $N_{\text{ens}}, \Delta t_m, t_{\text{trans}}, \Delta t, A_M/A_m, N_A, M_k$
\For{each continuum timestep}
  \For{each grid cell}
    \State Extract $T$, $\{pY_k\}$
    \For{$i = 1$ to $N_{\text{ens}}$}
      \State Run KMC, discard $t_{\text{trans}}$
      \State Record $\text{net}_k^{(i)}$
    \EndFor
    \State Avg: $\overline{\text{net}_k} = \frac{1}{N_{\text{ens}}} \sum \text{net}_k^{(i)}$
    \State Flux: $\frac{\partial Y_k}{\partial t} = \frac{\overline{\text{net}_k} A_M}{A_m N_A M_k \Delta t_m}$
    \State Apply flux to continuum
  \EndFor
  \State Advance continuum by $\Delta t$
\EndFor
\end{algorithmic}
\end{algorithm}
\end{minipage}%
\hfill
\begin{minipage}[t]{0.48\textwidth}
\begin{algorithm}[H]
\caption*{\textbf{(b) Surrogate-coupled (simplified)}}
\begin{algorithmic}[1]
\State \textbf{Input:} all of Algorithm 1, plus surrogate models and threshold $\tau_{\text{var}}$
\For{each continuum timestep}
  \For{each grid cell}
    \State Extract $T$, $\{pY_k\}$
    \State Query surrogate for variance $\sigma_k^2$
    \If{\textbf{any $\sigma_k^2 > \tau_{\text{var}}$}}
      \For{$i = 1$ to $N_{\text{ens}}$}
        \State Run KMC, discard $t_{\text{trans}}$
        \State Record $\text{net}_k^{(i)}$
      \EndFor
      \State Avg and compute flux (same as Algorithm 1)
      \State Retrain surrogate with new fluxes
    \EndIf
    \State Query surrogate for mean flux
    \State Apply flux to continuum
  \EndFor
  \State Advance continuum by $\Delta t$
\EndFor
\end{algorithmic}
\end{algorithm}
\end{minipage}
\end{tcolorbox}
\caption{Side-by-side comparison of directly coupled and surrogate-coupled algorithms for continuum–atomistic simulations. The directly coupled version (a) summarizes Algorithm~\ref{alg:directly_coupled}; the surrogate-coupled version (b) introduces only the differing logic for adaptively replacing KMC calls with surrogate queries.}
\label{fig:algo_comparison}
\end{figure}

A \gls{gp}-based surrogate is chosen since it has powerful predictive capability with sparse data and can provide robust uncertainty estimates. The latter will be crucial for quantifying when to construct additional on-the-fly training data as discussed below. An open-source Python-based software package named Adaptive Computing \cite{griffin2025adaptive} is used to implement surrogate modeling and active learning. As the simulation codes are written in C++, we utilize the official Python/C API, which enables us to call a python interpreter for the Adaptive Computing library inside the C++ continuum solver. The \gls{gp} implementation in the Surrogate Modeling Toolbox was adopted with default parameters \cite{SMT2019}. Specifically, a square exponential kernel was utilized, and the Truncated Newton-Conjugate Gradient method was employed to maximize the log-marginal likelihood. The optimization searched for the optimal kernel length scales within box constraints of [$10^{-6}$, 20], utilizing SMT's default multi-start strategy (10 random initializations) to avoid local optima.

Our adaptive computing framework creates a surrogate model that can in general take an arbitrary number of inputs and produce a single output. The surrogate model inputs are features of the continuum state that would affect atomistic processes. For the perovskite problem, this is the concentration of each of six ion species in the vapor jet, as well as the local temperature at the site of crystallization on the substrate. The output of a surrogate model is information relevant to the continuum state which is informed by atomistic processes. Deposition and desorption of each species onto a pervoskite crystal results in a net species flux for the continuum state. We thus employ six surrogate models, where each model predicts the net flux of one ion species onto the substrate (out of the domain) over one fluid timestep.

\subsection{Initial offline training}
Although the \gls{gp} is capable of learning and will conduct the substantial majority of its learning online (while the surrogate-coupled simulation is running), the \gls{gp} must be initialized with a small number of data points. To initialize the \gls{gp}, we perform a batch of \gls{kmc} simulations offline. The inputs of these atomistic simulations are determined by performing a CFD (non-surrogate) simulation of our domain and sampling temperature and species concentration at ten prescribed locations along the substrate at evenly spaced distances from the center of the jet. At each location, the same number of trials of \gls{kmc} simulations are performed as in the directly coupled simulations. As will be used for on-the-fly learning, the surrogate is trained on the average over these trials rather than the raw \gls{kmc} results so as to reduce the total number of training points (since the cost of training the \gls{gp} scales with the cube of the number of training points).

\subsection{Uncertainty-guided on-the-fly learning}
In our simulations, we will show that crystal growth (atomistic solution) steadily extracts each gas species from the continuum field, thereby producing a nearly-monotonic trend as opposed to an oscillatory or chaotic one. As this trend develops, the atomistic solver will be queried at novel input conditions that are not simply interpolations of the conditions covered in the offline training set nor previous online data points. 

Our surrogate model will therefore need to distinguish between cases in which interpolation of previous data suffices and cases where the model must learn more information by running an additional atomistic simulation. Simply put, the surrogate model should only use regression when confident in its prediction. The uncertainty (or inaccuracy) in the prediction is measured by the variance of the \gls{gp} output. We set a threshold value, which trusts the surrogate up to a specified variance value, and triggers a new atomistic simulation to improve accuracy when this threshold is exceeded. For the \gls{gp}, this variance accounts for the epistemic uncertainty of undersampling the underlying statistical process. This uncertainty tends towards zero as more samples are collected in the vicinity (in input parameter space) of some operating condition under consideration. This effectively means that atomistic simulations are performed in undersampled regions and the existing surrogate model is trusted in the vicinity of previous evaluations.

The continuum solver drives the execution. At each point on the substrate in the continuum solver, it queries the six adaptive computing surrogates for their estimates of the mean and variance. The highest variance returned is then compared to a fixed threshold. If the variance is below the threshold, the continuum solver deems the surrogate trustworthy and uses the surrogate's mean in place of calling the atomistic solver. Otherwise, the atomistic solver is called to calculate the six species fluxes and the surrogate models are retrained with this new atomistic solver solution.

\section{Results} \label{sec:results}
We evaluate the accuracy of our Adaptive Computing framework for simulations of vapor flowing onto a reactive substrate with consideration of atomistic effects from perovskite crystal formation. We consider both the directly-coupled and surrogate-modeled approaches outlined in Figure~\ref{fig:summary}(a) and (c), respectively. In the directly-coupled simulation, every grid point along the substrate employs the \gls{kmc} atomistic solver to calculate fluxes of ion species, updating the continuum solver with this information as detailed in Algorithm~\ref{alg:directly_coupled}. In surrogate-coupled simulations, every grid point along the substrate employs surrogate models that learn on-the-fly. We explore three surrogate-coupled simulations with different tolerances of uncertainty.

\begin{figure}
    \centering
    \includegraphics[width=0.7\linewidth]{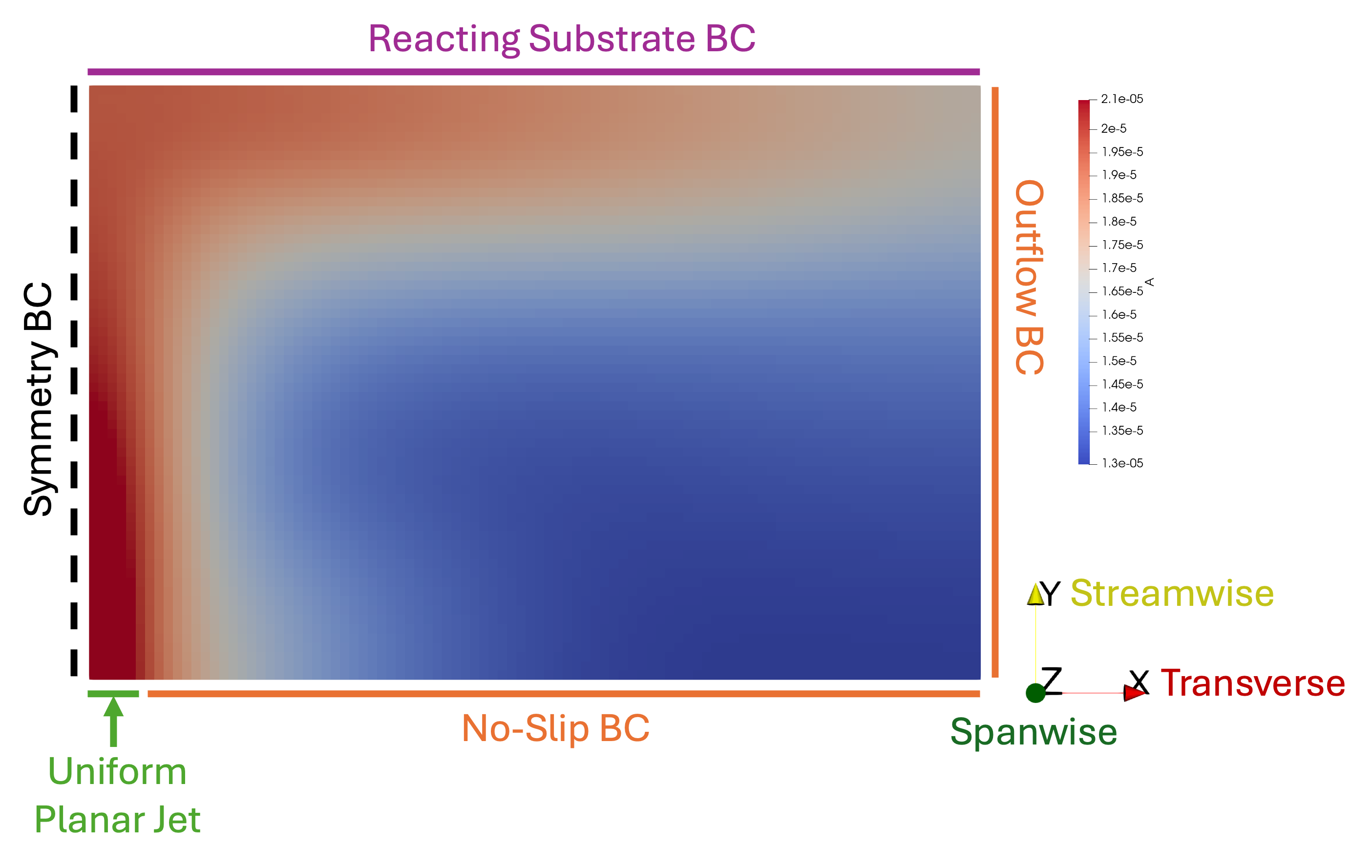}
    \caption{Schematic diagram of a cross section of the simulation domain and boundary conditions (BCs) for a planar jet impinging on a reacting substrate, color indicates the partial pressure of species A in bars. Not pictured are the periodic boundary conditions in the out-of-plane spanwise direction.}
    \label{fig:boundary_conditions}
\end{figure}

A quasi-2D planar jet (flow issuing from an infinite slot) impinging on a reactive substrate is simulated in a rectangular 3D Cartesian domain, with $x$, $y$, and $z$ axes corresponding to the transverse, streamwise (jet inflow direction), and spanwise (periodic) dimensions as shown in Fig. \ref{fig:boundary_conditions} with resolution $96 \times 64 \times 2$, respectively. This means our domain is 15~cm in the transverse direction, 15 times wider than the width of the incoming jet. Other works have found 10-20 jet widths to be sufficiently wide \cite{secchi2026energetics, nithyanandham2026nozzle, neumann2011grid}. As presented in Fig.~\ref{fig:boundary_conditions}, there is not significant interaction with the wall, as seen through the monotonic decrease in partial pressure of the jet material along the transverse axis.

Each cell is a cube of side length 1.56~mm. The uniform jet issues in the +y direction from the section of the boundary marked in green in Fig. \ref{fig:boundary_conditions}. The transverse centerline uses a symmetry boundary condition, with the normal velocity reflect-odd and all other quantities reflect-even, corresponding to $u_i=0$ and $\partial q/\partial x_i=0$. The outer transverse boundary uses first-order extrapolation, meaning $q_g=q_b$ for all ghost cells $q_g$ and the boundary cell $q_b$. The spanwise direction is periodic. The streamwise upper and lower walls use external-Dirichlet conditions; outside the inlet the velocity is reflect-odd (no-slip) at the wall, while the inlet patch imposes $\mathbf{u}=(0,U_{\rm in},0)$, $T=T_{\rm in}$, and $Y_i=Y_{i,\rm in}$.

\begin{table}[]
\centering
\begin{tabular}{l|c|c|c|c|c|c}
    species & A ($\mathrm{Cs^+}$) & B  ($\mathrm{Pb^{2+}}$) & X  ($\mathrm{I^-}$) & AX ($\mathrm{CsI}$) & BX ($\mathrm{PbI^+}$) & BX$_2$ ($\mathrm{PbI_2}$) \\
    \hline
    initial $Y_k$ & 0.0015 & 0.0015 & 0.002 & 0.002 & 0.002 & 0.001
\end{tabular}
\caption{Initial mass fraction $Y_k$ of the planar jet. The remaining mass is \ce{N2}.}
\label{tab:massfrac}
\end{table}

Simulations begin with a box of \ce{N2} with pressure of 10 Torr (1333.3~Pa) and density 0.018~kg/m$^3$. In the first timestep, a jet is released from an inlet 5~mm in half-width (symmetry condition implies an effective full width of 10~mm). Table~\ref{tab:massfrac} provides the jet composition, and unless otherwise specified, the jet has a flow speed of 50~m/s at the point of entry. An equilibrium state with a non-reacting substrate is achieved over about 0.8~s. This ensures that the impinging jet becomes steady and achieves a stagnation flow profile as shown in Fig.\ \ref{fig:boundary_conditions}. Reacting simulations are then conducted over a total time 0.39~ms that was sufficient to achieve steady growth rate.

Both the directly-coupled and surrogate-coupled simulations employ KMC atomistic simulations. These consider a 3D subdomain consisting of a doubly periodic horizontal patch on the substrate of size $N_x=10$ and $N_y=10$ (87$\times$87~\AA), along with $N_z=20$ grid cells (174~\AA) for a crystal to grow vertically. This subdomain represents a very small component of one fluid grid cell along the substrate in the continuum simulation. Net deposition onto the pervoskite crystal is calculated in the subdomain, and a species flux out of (or into) the continuum simulation is obtained by multiplying this deposition by the ratio of the continuum grid area to the subdomain area. 

The purpose of the \gls{kmc} simulation is to evaluate a rate of net deposition for each species.
To calculate this rate, each fluid timestep requiring an atomistic simulation (i.e., \emph{not} well approximated by the surrogate) triggers a \gls{kmc} run lasting $10\,\mu\mathrm{s}$. Note that this simulation window is intentionally chosen to be 100 times longer than a fluid timestep, which enables us to achieve more converged statistics of average deposition rate and reduce the stochasticity associated with smaller averaging windows.

\begin{figure}
    \centering
    \includegraphics[width=0.9\linewidth]{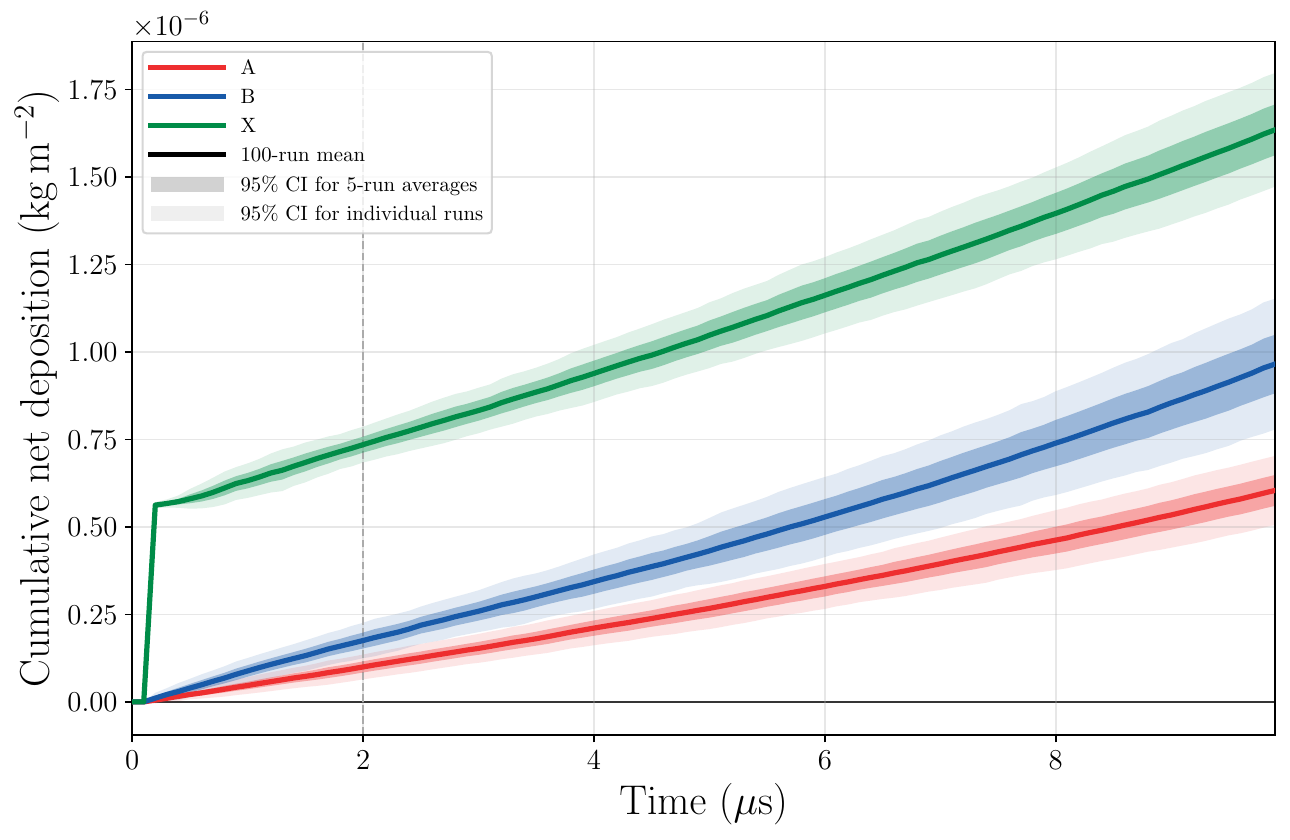}
    \caption{Cumulative net deposition of species A, B, and X is presented as a function of time. We sample one hundred KMC simulations with consistent partial pressures \(P_A = 7.5\times10^{-5}\,\mathrm{bar}\), \(P_B = 7.5\times10^{-5}\,\mathrm{bar}\), \(P_X = 1.0\times10^{-4}\,\mathrm{bar}\), \(P_{AX} = 1.0\times10^{-4}\,\mathrm{bar}\), \(P_{BX} = 1.0\times10^{-4}\,\mathrm{bar}\), \(P_{BX2} = 5.0\times10^{-5}\,\mathrm{bar}\), and temperature 300~K. Solid curves show the simulated cumulative net deposition, and shading shows corresponding confidence intervals. The important feature is that sampling 5-run averages reduces variance and that the curves from \(2\text{--}10\,\mu\mathrm{s}\) are mostly linear, corresponding to a constant deposition rate.}
    \label{fig:transient}
\end{figure}

Each KMC solve is initialized with a clean substrate since the simulations are temporally decoupled from the continuum solve and there is no obvious way to initialize the crystal state to be representative of a crystal growing steadily at the current instantaneous continuum conditions. This leads to an initial transient stage in each atomistic simulation wherein crystal growth from a blank substrate expresses a stochastic net deposition rate that is unrepresentative of long-time behavior. Following this transient stage, an average is taken over the subsequent period when a relatively time-invariant net deposition rate is observed (Figure~\ref{fig:transient}). 

The emergence of time invariant crystal growth rates is related to the emergence of a fully developed wetted surface of the crystal that leads to similar growth regardless of the number of occupied crystal layers beneath the wetted surface. Data from this initial stage, with transient time $t_\mathrm{trans}$, are discarded in our calculations because in a real experimental system, a continuously growing crystal initialized from a clean substrate would spend a negligible fraction of its time in this transient phase compared to the relatively long timescale of crystal development. Our simulations take $t_\mathrm{trans}=2\,\mu\mathrm{s}$, though greater accuracy may be achieved by increasing the transient interval (which proportionately increases the computational cost of each KMC simulation). At each continuum grid cell and for each fluid timestep, an ensemble of $N_\mathrm{ens}=5$ \gls{kmc} simulations is performed with the resulting deposition rates averaged. This reduces the inherent stochasticity in our KMC estimates by a little over half compared to averaging over a single simulation, though with a greater computational budget, one could reduce variance further by a factor of $\sqrt{N_\mathrm{ens}}$. The remaining stochasticity that enters the surrogate- and direct-coupled simulations is quantified by the narrower set of confidence intervals in Fig.~\ref{fig:transient}.

For all simulations we track the total deposition on the substrate and the impact that this mass transfer has on the partial pressures of the six transported species.



\subsection{Surrogate-coupled simulations}
We consider three surrogate-coupled simulations, each using a different uncertainty threshold (above which the surrogate is considered too uncertain, necessitating a full KMC simulation). The threshold limits the variance of the \gls{gp} prediction measured in molecules deposited within the patch of one \gls{kmc} simulation ($10 \times 10$ cells or $80\times 90$ \AA). With on-the-fly learning, the initially naive surrogate model will defer to atomistic simulations more often. Upon learning from each new atomistic simulation, the uncertainty in surrogate predictions will more regularly measure below the set threshold, reducing the need to employ atomistic simulations.

\begin{figure}
\centering

\begin{overpic}[width=0.48\linewidth]{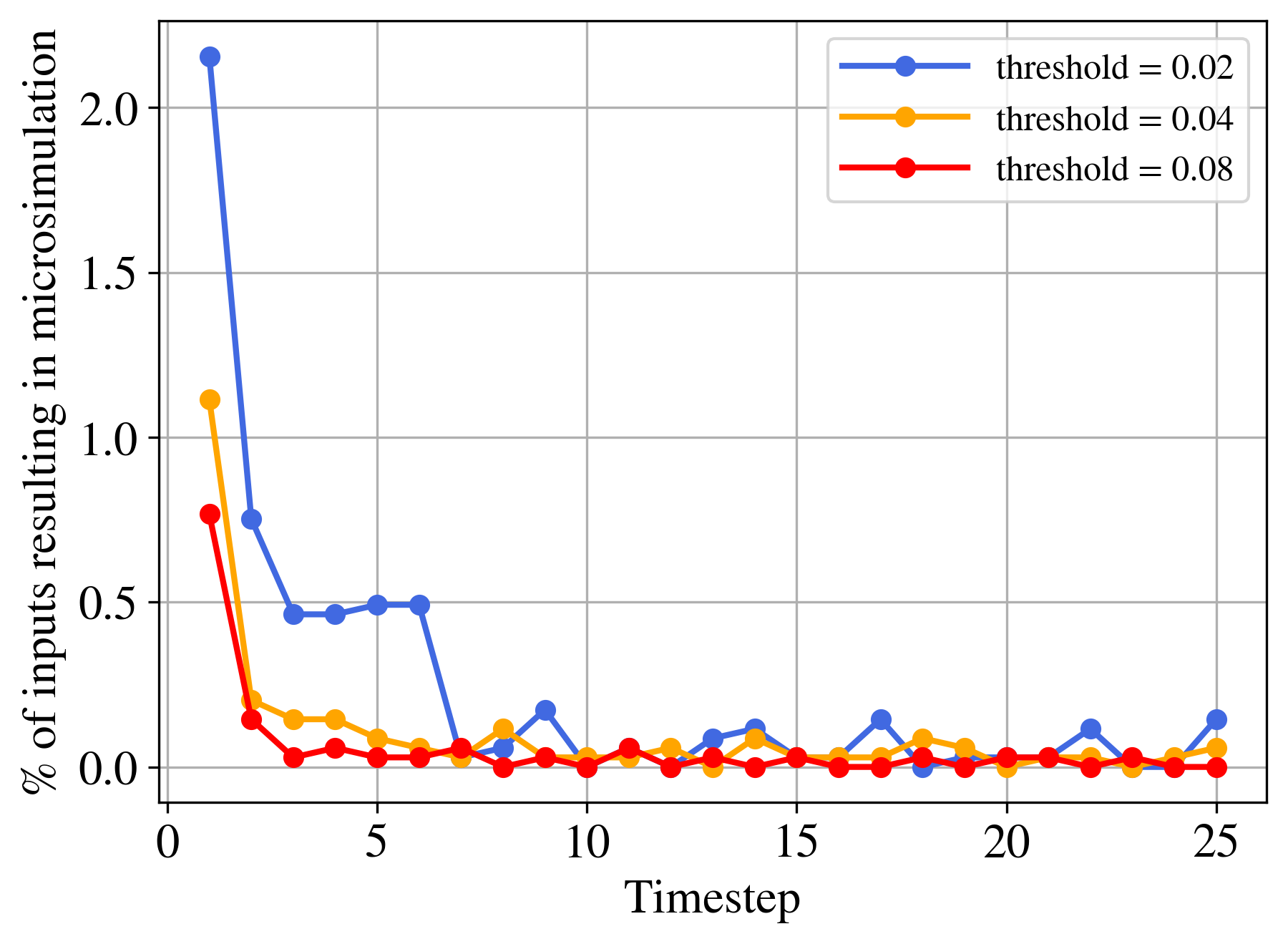}
\put(2,75){\textbf{(a)}}
\end{overpic}
\hfill
\begin{overpic}[width=0.48\linewidth]{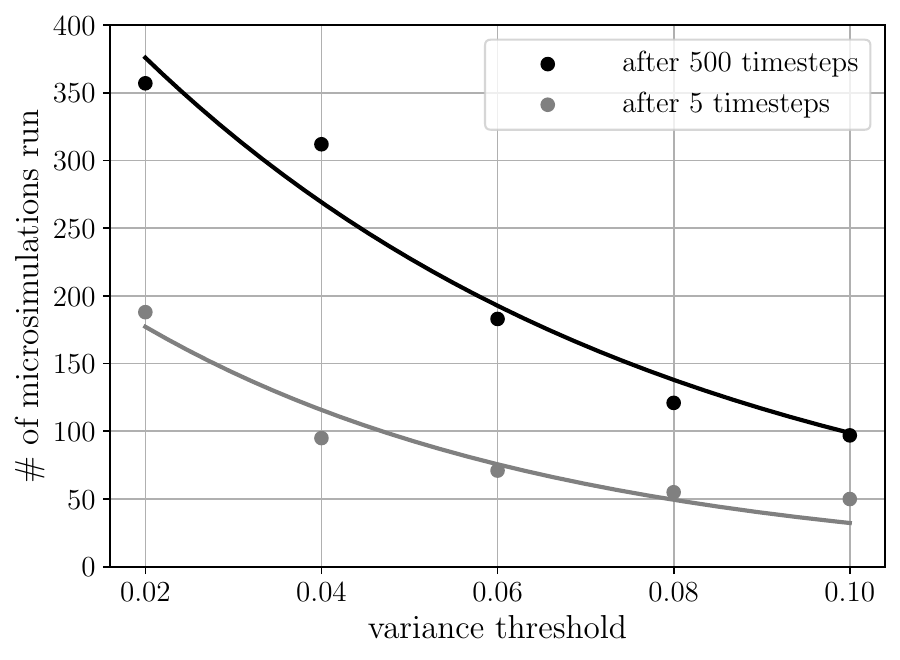}
\put(2,75){\textbf{(b)}}
\end{overpic}

\caption{\textbf{(a)} For three surrogate models each using an uncertainty threshold of 0.02 (blue), 0.04 (yellow), and 0.08 (red), we present the percent of grid cells that require performing a high-fidelity atomistic simulation as opposed to relying on the surrogate output. \textbf{(b)} The total number of atomistic simulations run after 5 timesteps (gray) and 500 timesteps (black) is presented as a function of variance threshold. The solid curves are logarithmic curves of best fit.}
\label{fig:HFcount}
\end{figure}

Figure~\ref{fig:HFcount} discusses the runtime benefits of employing the surrogate model. Due to potential differences in computing architecture, runtime is discussed by using the number of atomistic simulations run as a proxy. The runtime of directly-coupled simulations $R_\mathrm{dc}$ scales linearly with the corresponding number of atomistic simulations $N_\mathrm{dc}$ that the total simulation requires, i.e., $R_\mathrm{dc} \sim N_\mathrm{dc}$. 
In the surrogate-coupled simulations, the number of atomistic simulations required $N_{sc}$ is less than in the directly coupled case, but there are two contributions to the runtime cost: a linear contribution just as before associated with running the simulations and a cubic contribution that accounts for the cost of evaluating the surrogate model. This cubic term arises because Gaussian process regression requires factorizing or inverting an $N_{sc} \times N_{sc}$ covariance matrix, a standard $\mathcal{O}(N_{sc}^3)$ operation; here, each atomistic simulation supplies one training point, so the GP training-set size grows directly with $N_{sc}$. Thus, the runtime of surrogate-coupled simulations $R_\mathrm{sc} \sim N_\mathrm{sc} + C N_\mathrm{sc}^3$, where $C$ is a small constant that is approximately the relative cost of an atomistic simulation to evaluating the surrogate model when trained on only a single atomistic simulation. This model therefore combines the linear cost of additional atomistic evaluations with the cubic scaling inherent to GP updates. Since $C$ is small, for small values of $N_\mathrm{sc}$, the linear term dominates. Eventually, the faster growing cubic term dominates, but $R_\mathrm{sc}$ remains less than $R_\mathrm{dc}$. Finally, the cubic term is large enough that even though $N_\mathrm{sc} << N_\mathrm{dc}$, the cost of a surrogate simulation exceeds that of a directly coupled simulation. 

In practice, the case in which surrogate simulation exceeds direct simulation in cost should be avoided through the selection of an appropriate type of surrogate model. GPs are known to perform well for ``sparse data" problems and scale poorly for ``big data" applications. Other methods such as mixture density networks or Bayesian neural networks should be deployed in this ``big data" regime. Conversely, for ``sparse data" applications in the linear regime, the GP will out perform these other approaches by given more accurate predictions at comparable computational expense. Thus, when a GP is used, there is a practical variance threshold limit after which the cubic term leads to prohibitively expensive $R_\mathrm{sc}$. To predict speedup and select an appropriate variance threshold for one's surrogate model, a reference runtime proportion may be obtained by comparing the directly-coupled simulation against a surrogate-coupled simulation for one choice of variance threshold. Then, the choice of variance threshold may be informed by the logarithmic relation of threshold to number of atomistic simulations (Fig.~\ref{fig:HFcount}b) and the nearly cubic scaling of number of atomistic simulations to runtime.

Figure~\ref{fig:HFcount}a verifies that the number of atomistic simulations needed reduces drastically after sufficiently many timesteps. The number of atomistic simulations in a directly coupled simulation is the number of surface grid points times the number of time steps, i.e., 192 per timestep, 960 after 5 time steps, and 96,000 after 500 time steps. Even in first timestep, an atomistic simulation is only executed at roughly one percent of grid cells (with a lower uncertainty threshold requiring more atomistic simulations). Across the variance thresholds explored, within the first ten timesteps, the frequency with which an atomistic simulation is called soon approaches a final state, hovering near zero with some timesteps requiring no atomistic simulations and some requiring an atomistic simulation $\mathcal{O}(0.1\%)$ of the time (for our domain setup, this means just a few atomistic simulations). Though not presented in the figure, this state persists long term through one-thousand timesteps. 

The choice of variance threshold has some influence on the long-term state, where the lowest-threshold (0.02) case exhibits fewer timesteps with zero atomistic simulations. A larger influence of the choice of threshold is observed in the first several timesteps, where the lowest-threshold case sustains a non-negligible proportion of atomistic simulations for multiple timesteps, while the other cases decay to an $\mathcal{O}(0.1\%)$ proportion of atomistic simulations after the first timestep. 

This behavior suggests a practical upper and lower bound on the choice of threshold. Figure~\ref{fig:HFcount}b shows that the number of required atomistic simulations scales logarithmically with variance threshold. Beyond a certain threshold, one risks reducing accuracy for rapidly increasing computational expense. A reduction in variance threshold leads to a super-linear increase in number of atomistic simulations and computational cost scales near-cubically with this number. The gray curve in Figure~\ref{fig:HFcount}b shows that the number of atomistic simulations approaches nearly half of its eventual value within the first five timesteps, meaning that computational cost (and therefore runtime-per-timestep for all subsequent timesteps) increases sharply at the beginning before leveling out for subsequent timesteps. This scaling of computational cost implies there exists a practical variance limit beyond which the surrogate modeling becomes more expensive than the directly coupled approach. 

Further reasoning behind the lower bound on variance threshold comes from a reduction in accuracy. Not presented in Figure~\ref{fig:HFcount}, for the same model parameters we ran one simulation with a variance threshold of 0.01. In addition to the ramp up in runtime, we observed unreasonable surrogate predictions, likely due to overfitting of the \gls{gp} once the variance threshold undershoots the natural variance of the \gls{kmc} simulations.

\begin{figure}
    \centering
    \includegraphics[width=1.0\linewidth]{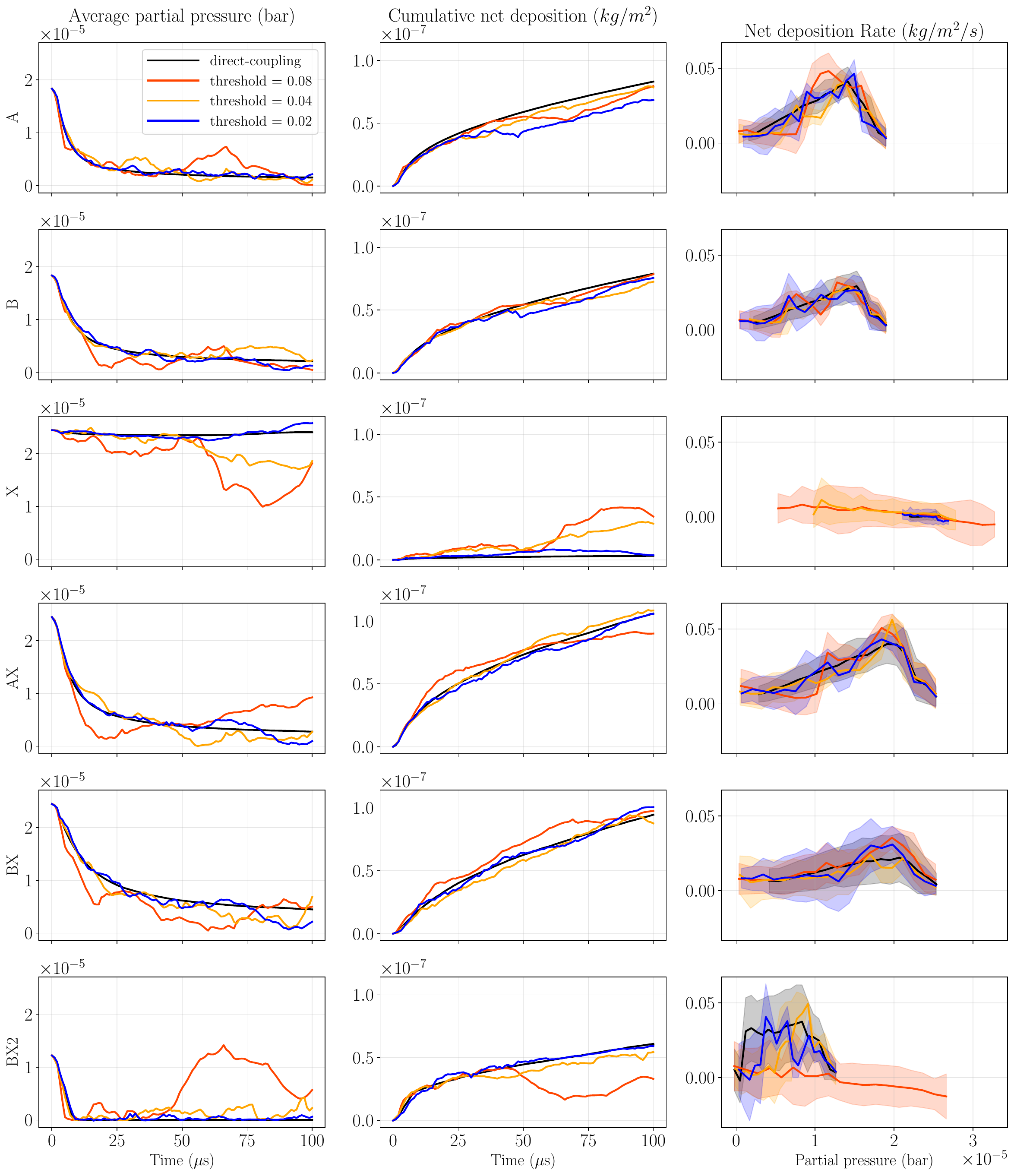}
    \caption{Continuum species concentrations, culmulative net depositions, and net deposition rates as predicted by our surrogate model at three different variance thresholds (0.02 in blue, 0.04 in yellow, 0.08 in red) are compared against the directly-coupled model (in black). Rows, from top to bottom, correspond to species A, B, X, AX, BX, and BX$_2$ (A is $Cs$, B is $Pb$, and X is $I$). Columns present (\textbf{left}) the partial pressure (bar) along the boundary, (\textbf{middle}) cumulative deposition along the substrate (kg/m$^2$), and (\textbf{right}) the learned net deposition rate (kg/m$^2$/s) as a function of partial pressure. For the latter, the solid line is the average calculated rate, with shading showing one standard deviation.}
    \label{fig:accuracy}
\end{figure}

\begin{figure}[!htbp]
    \centering
    \includegraphics[width=1.0\linewidth]{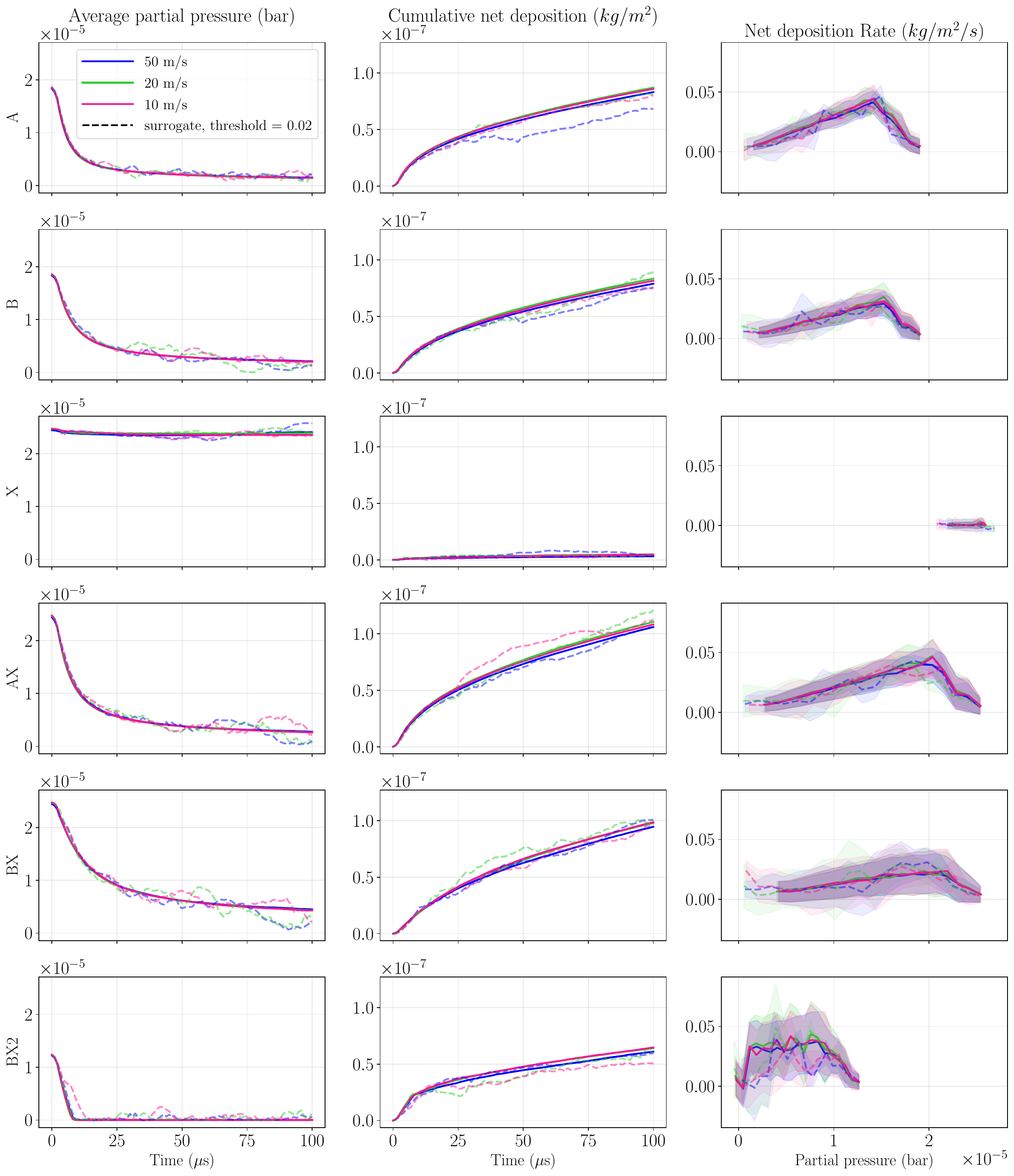}
    \caption{The same data as outlined in Fig.~\ref{fig:accuracy} is presented for simulations of various jet flow speeds. Solid lines correspond to a directly-coupled simulation while dashed lines are surrogate-coupled simulations all using a variance threshold of 0.02.}
    \label{fig:flowcomparison}
\end{figure}

Though computational demand increases with a lower variance threshold, so should the accuracy of the surrogate model. Figure~\ref{fig:accuracy} compares model outputs for three surrogate-coupled simulations (of different variance threshold) against a directly-coupled simulation, which is treated as the ground truth. 

Before comparing the accuracy dependence on variance threshold, we first discuss the broad behavior of the directly-coupled model we wish for the surrogates to capture. The left and middle columns show over time how the partial pressure of each species in the domain generally decreases while the cumulative deposition of each species onto the substrate increases. The right column plots the variables that the surrogate model is formulated in terms of, that is, the net deposition rate versus the partial pressure. In the \gls{kmc} simulations, the deposition of each individual species on the substrate is a function of temperature and the partial pressures of every species. The atomistic simulations calculate this dependence, while the surrogate models learn this dependence empirically. The right column shows, averaging over every timestep and grid cell, the resulting deposition rate for a species plotted over the local partial pressure of that species (for the sake of visualization, the other dependent variables are disregarded but influence the variation in the resulting curve).

The right column of Figure~\ref{fig:accuracy} shows that for each species, the deposition rate increases with partial pressure of that species up to a critical value after which for higher pressures, the deposition rate will actually decrease (the exception being species X which has a consistent near-zero deposition rate). The former trend reflects the intuition that when a species is in higher abundance, there is more available material to deposit. The latter trend is initially unintuitive and may indicate that oversaturation inhibits deposition or that oversaturation is correlated with other conditions such as desorption events or a change in temperature that inhibits deposition. The time-dependent curves in the left and middle columns show that after the first 10~$\mu$s or so, the domain is in a relatively low-pressure environment. As expected, for all models presented, the downswings or upswings in partial pressure (left column) are indeed associated with upswings or downswings in cumulative deposition (middle column) respectively. 

The directly-coupled simulation suggests a smooth reduction in partial pressure of each species down to some equilibrium value, along with a smooth increase in cumulative net deposition eventually approaching a linear trend with constant deposition rate. The surrogate models capture this behavior, with deviations presenting mostly as oscillations around the smooth directly-coupled prediction. These oscillations tend to increase with the variance threshold of the surrogate model. Notably in the cases of species X and BX, we observe the the higher-variance surrogate models cannot maintain a domain of near-unchanging pressure, wrongly predicting a downswing in the pressure of X and upswing in BX$_2$, made more severe in the 0.08-threshold case relative to the 0.04-threshold case.  In the other species (A, B, AX, BX), we likewise observe greater oscillation in the pressure curve for the 0.08-threshold case, while the 0.04 and 0.02-threshold cases agree more closely. We note, however, the oscillations in pressure for the 0.08-threshold case do not prevent the model from predicting a reasonable estimate of net deposition for the A, B, AX, and BX cases. 

Regarding the learned pressure-dependence of deposition rate presented in the right column of Figure~\ref{fig:accuracy}, all surrogate models capture the appropriate trend (direct proportionality up to a critical pressure followed by inverse proportionality) for each species except for X and BX$_2$. For species X, the higher-variance surrogate models overestimate the deposition rate, which is consistent with the inaccurate upswing in cumulative deposition at later times. For species BX$_2$, the higher-variance surrogate models underestimate the deposition rate, which is consistent with the desorption at later times.

In addition to the above simulations, Figure~\ref{fig:flowcomparison} considers variation in the jet flow speed, where the solid curves correspond to directly-coupled simulations and the dashed curved are surrogate-coupled simulations. Recall simulations are run under a non-reacting substrate (an uncoupled purely fluid dynamical process) to a spun-up solution before the \gls{kmc} simulations are coupled. The 50~m/s case discussed above is considered here as well as velocities of 20~m/s and 10~m/s. We accordingly observe in Figure~\ref{fig:flowcomparison} that all coupled simulations start from nearly the same initial conditions. As such, the directly-coupled simulations also advance nearly identically, with slight variation in the cumulative net deposition at long time. 

The surrogate-coupled simulations in Figure~\ref{fig:flowcomparison} all use the same variance threshold of 0.02. The accuracy of the low-threshold surrogates is consistent across jet flow speeds with no discernible pattern in the dependence of inaccuracy on flow speed. We conclude the surrogate modeling approach is robust over variations in the flow speed of the jet.

\begin{table}
\centering
\begin{tabular}{cc|cccccc}
\hline
Variance threshold & Flow speed & A & B & X & AX & BX & BX2 \\
\hline
0.02 & 10~m/s
& \cellcolor{accgreen}93.52\%
& \cellcolor{accgreen}92.05\%
& \cellcolor{accdarkgreen}97.55\%
& \cellcolor{accdarkgreen}96.10\%
& \cellcolor{accdarkgreen}98.45\%
& \cellcolor{accgreenyellow}78.25\% \\

0.02 & 20~m/s
& \cellcolor{accgreen}93.52\%
& \cellcolor{accgreen}93.01\%
& \cellcolor{accdarkgreen}99.35\%
& \cellcolor{accgreen}90.51\%
& \cellcolor{accdarkgreen}97.88\%
& \cellcolor{accgreen}93.66\% \\

0.02 & 50~m/s
& \cellcolor{accgreenyellow}82.36\%
& \cellcolor{accdarkgreen}95.83\%
& \cellcolor{accgreenyellow}83.47\%
& \cellcolor{accdarkgreen}99.67\%
& \cellcolor{accgreen}93.56\%
& \cellcolor{accdarkgreen}97.11\% \\

0.04 & 50~m/s
& \cellcolor{accgreen}94.49\%
& \cellcolor{accgreen}91.93\%
& \cellcolor{accred}-706.11\%
& \cellcolor{accdarkgreen}97.72\%
& \cellcolor{accgreen}92.77\%
& \cellcolor{accgreenyellow}89.33\% \\

0.08 & 50~m/s
& \cellcolor{accdarkgreen}95.59\%
& \cellcolor{accdarkgreen}99.47\%
& \cellcolor{accred}-883.89\%
& \cellcolor{accgreenyellow}84.98\%
& \cellcolor{accdarkgreen}96.81\%
& \cellcolor{accyellow}54.32\% \\
\hline
\end{tabular}
\caption{Accuracy (Eq. \ref{eq:accuracy}) (100 - percent error) of the final cumulative net deposition for the surrogate models presented in Figures \ref{fig:accuracy} and \ref{fig:flowcomparison} relative to their corresponding directly-coupled models.}
\label{tab:cumulative_deposition_accuracy}
\end{table}

Table~\ref{tab:cumulative_deposition_accuracy} summarizes the performance of the surrogate models presented. The final value of cumulative net deposition for each gas species is compared between the directly-coupled simulations and their corresponding surrogate models, where accuracy is calculated as
\begin{equation}
\text{Accuracy} =
\left(
1 -
\frac{
\left|D_{\mathrm{surrogate}} - D_{\mathrm{direct}}\right|
}{
\left|D_{\mathrm{direct}}\right|
}
\right)
\times 100\%,
\label{eq:accuracy}
\end{equation}
where $D_{\mathrm{surrogate}}$ and $D_{\mathrm{direct}}$ are the final values of cumulative net deposition for the surrogate-coupled and direct-coupled models respectively. Because the final cumulative net deposition is equivalently the time-integrated deposition rate of the simulation, this error metric also measures the time-averaged error in the simulation's net deposition rate. As expected, the percent errors are larger when the GP's variance threshold is larger. That is, a higher variance threshold implies that the GP predicts that surrogate-coupled results have a higher degree of epistemic uncertainty, and the higher percent error indicates that indeed the surrogate-coupled simulations deviate more from the directly coupled simulations. The accuracy for species X is particularly low for variance thresholds of 0.04 and 0.08. This is explained by the relatively low magnitude of $D_{\mathrm{direct}}$, the directly coupled net deposition rate, for species X compared to this value for other species as shown in Fig.~\ref{fig:accuracy}. Specifically, similar absolute error exists for X and BX2 at a variance threshold of 0.08, but the relatively low absolute value of $D_{\mathrm{direct}}$ for species X leads to a correspondingly high percent error. The GP training process maximizes the marginal log-likelihood of the GP length scale (fitting) parameters given the observations, and there is no guarantee that relative errors will be consistent across species. The key observation is that the errors generally decay as the variance threshold is reduced. The surrogate-coupled models with the lowest variance threshold (0.02) perform with at least 78\% accuracy across all species and across all flow speeds (with the case of 20~m/s flow admitting \>90\% accuracy across all species). Across all surrogate-coupled models presented, the final cumulative net deposition species B and BX is accurately predicted within 91\%, though species X and BX2 show more severe inaccuracies when the variance threshold reaches 0.04 and above.

In Appendix \ref{app:constant_flux}, we explore if a simple constant flux model can be used in place of the GP surrogate model. It is observed that the surrogate-coupled simulations agree significantly better with directly coupled simulations than the simulations based on the constant flux assumption.


\section{Conclusions} \label{sec:conclusions}

In this work, we presented an autonomous, uncertainty‑aware surrogate modeling framework for multiscale coupling between atomistic surface‑chemistry simulations and continuum reactive‑flow models. The approach replaces most direct KMC evaluations with online‑trained surrogate model evaluations that provide species‑resolved boundary mass fluxes with quantified uncertainties. The continuum solver uses these uncertainties to trigger additional KMC simulations only when needed, enabling accurate and efficient two‑way coupling across disparate spatial and temporal scales.

The domain problem presented considers vapor deposition onto a substrate for perovskite crystal formation. The framework combines a multi‑species compressible CFD solver with an atomistic KMC surface‑chemistry model, which incorporates a newly developed DFT‑calibrated energetics model. Surrogate models trained on‑the‑fly reduce the frequency with which expensive atomistic simulations must be performed. By comparing directly‑coupled and surrogate‑coupled simulations, we assessed both the computational efficiency and predictive accuracy of the Adaptive Computing approach. Surrogate‑coupled simulations reproduce key physical trends observed in the directly coupled model, including the non‑monotonic relationships between deposition rate and partial pressure across species. For a conservative uncertainty threshold, cumulative mass‑flux errors typically fall within 10–20\%, while reducing the required number of KMC simulations by two to three orders of magnitude. This comparison demonstrates that the surrogate‑coupled approach substantially reduces the number of required atomistic simulations without sacrificing the overall physical behavior captured by the directly‑coupled approach. After an initial learning stage spanning only 5–10 fluid timesteps, the surrogate model is able to provide predictions for nearly all grid cells along the substrate, with atomistic simulations being required only sporadically thereafter. This implies that the majority of computational cost occurs upfront; extending simulations well beyond the learning stage is extremely inexpensive. The number of atomistic simulations required exhibits near‑logarithmic dependence on the chosen variance threshold for the surrogate model.

The variance threshold thus provides a practical mechanism for balancing computational cost and predictive accuracy. Lower thresholds increase the number of atomistic simulations required during the early stages of the simulation, while higher thresholds reduce this cost but introduce larger oscillations in predicted quantities. Across the range of thresholds examined, surrogate‑coupled simulations were able to reproduce the qualitative trends of the directly‑coupled simulation, including the smooth reduction of partial pressures over time and the eventual establishment of steady deposition rates. However, higher variance thresholds led to greater variability in the deposition trend for species X and BX$_2$.

The surrogate models successfully learned the fundamental dependence of deposition rate on local partial pressure across most species. In particular, the models captured the non‑monotonic behavior observed in the directly‑coupled simulations, wherein deposition rates increase with partial pressure up to a critical value before decreasing at higher pressures. Surrogate model accuracy using a low variance threshold (0.02) is robust across variations in jet flow speed. This agreement indicates that the surrogate models are capable of learning physically meaningful relationships directly from atomistic simulation data during the course of a simulation.

Importantly, our results demonstrate that for this multiscale surface‑chemistry application, surrogate models need not represent the evolving crystal structure explicitly to provide accurate atomistic‑to‑continuum coupling. The six species‑resolved mass fluxes constitute a sufficiently low‑dimensional representation of the atomistic response, enabling substantial computational savings without a significant compromise in accuracy.

Our surrogate‑coupled model of gas flow and perovskite growth was built using our open‑source Adaptive Computing code that can be flexibly applied to a wide range of modeling problems. We have used the GP surrogate model, which is ideally suited for accurate predictions with sparse training data. If significantly higher accuracy (at the cost of more frequent KMC simulations) is desired, neural‑network‑based methods with quantified uncertainty, such as mixture‑density networks and Bayesian neural networks, could be employed. Future extensions may include exploring these neural‑network‑based surrogates, integrating adaptive cost-aware uncertainty thresholds \cite{Domantay2024}, and applying the approach to other reactive surface‑growth processes. On the experimental side, validation of the predicted deposition behavior against laboratory measurements may follow the approach demonstrated by Hassanaly et al. \cite{hassanaly2021surface}.

\backmatter

\bmhead{Supplementary information}

Tables with computed adsorption and desorption energies and details on the energy model used for the Kinetic Monte Carlo simulations are provided in Online Resource 1 (ESM\_1.pdf). Initial positions before relaxation for the different adsorption and defect calculations are provided as a zip file of VASP input files in Online Resource 2  (ESM\_2.zip).

\bmhead{Acknowledgments}

This work was authored in part by the National Laboratory of the Rockies (NLR) for the U.S. Department of Energy (DOE) under Contract No. DE-AC36-08GO28308. This work was supported by the Laboratory Directed Research and Development (LDRD) Program at NLR. The views expressed in the article do not necessarily represent the views of the DOE or the U.S. Government. The U.S. Government retains and the publisher, by accepting the article for publication, acknowledges that the U.S. Government retains a nonexclusive, paid-up, irrevocable, worldwide license to publish or reproduce the published form of this work, or allow others to do so, for U.S. Government purposes. A portion of the research was performed using computational resources sponsored by the Department of Energy’s Office of Critical Minerals and Energy Innovation and located at the National Laboratory of the Rockies. D.A. acknowledges support from the U.S. Department of Energy, Office of Science, Office of Advanced Scientific Computing Research, Department of Energy Computational Science Graduate Fellowship Award Number (DE-SC0023112).

\section*{Declaration of Competing Interests}
The authors declare no competing interests.

\begin{appendices}

\section{KMC computational details}\label{app:KMC}

\subsection{Event types and rates}

In this sub-section we describe each type of event and its associated rate in our KMC model. All energies (site, desorption and barrier energies) used for the KMC model are taken from a neighbor-based energy model that is described in the Appendix \ref{app:energy_model}.

\noindent Notation: $k_B$ denotes Boltzmann's constant; $T$ is the absolute temperature of the substrate; $m_\alpha$ is the mass of a species denote $\alpha$ (e.g., $A$ or $BX^+$); $p_\alpha$ is the partial pressure of species $\alpha$ in the gas phase directly above the surface; 

\begin{enumerate}

\item \emph{Deposition of the following species: A$^+$, B$^{2+}$, X$^-$, AX, (BX)$^+$, BX$_2$.}

The rate of deposition of each species, denoted $\alpha$, is given by the rate of impacts per unit area from the kinetic theory of gases multiplied the area represented by a site, $a_{site}$, multiplied by a sticking coefficient, $\phi_\alpha$: $R_{deposition}^\alpha = a_{site}\phi_\alpha p_\alpha/\sqrt{2\pi k_B T m_\alpha}$. For simplicity, in our simulations the sticking coefficients are set to 1.  To estimate deposition rates, we average over incident angle to compute the average rate of impingement per area times the area of a site. For molecular depositions onto a given lattice site, decisions need to be made about which atoms stick at the chosen site and which go to neighboring sites. We have implemented various options (statistical branching, lowest energy configuration, and probabilistic branching with Boltzmann-weighting of different adsorption configurations based on the energy). For the simulations described here, deposition rates are modified from the gas-phase rate per site based on the energy of the depositing species at each site according to the following rules. First when a species from the gas phase would have a positive energy compared to the gas phase (i.e., a negative adsorption energy) at a surface site $i$, the rate of deposition of that species at $i$ is set to zero. Second, when a species can deposit at a site in multiple positive-adsorption-energy configurations (e.g., an AX molecule could deposit A at site $i$ and $X$ at site $i+1$, or vice versa), we used a probabilistic Boltzmann weighting method to modify the gas-phase deposition rate of species $\alpha$ at site $i$ for each possible configuration, $\gamma$,
\begin{equation}
R_i^{\alpha,\gamma} = R_{deposition\ at\ i}^\alpha \frac{e^{-E_i^\alpha(\gamma)/k_B T}}{\sum_{\gamma^\prime} e^{-E_i^\alpha(\gamma)/k_B T}} ,
\label{bolzmann_weighting}
\end{equation}
where $\gamma$ and the sum $\gamma^\prime$ include only configurations with a positive adsorption energy. The idea behind this weighting approach is that a molecule that collides with a surface can over short times rattle around and sample a quasi-equilibrium distribution for a species bound to a particular site. 

\item \emph{Desorption from the surface.}

Any of the species in the model also can desorb. The rate is based on transition state theory, $R_{desorption}^{\alpha} = \nu_\alpha e^{-E_{desorb}/k_B T}$. $E_{desorb}^\alpha$ is the energy change in the system that happens when the species is moved from the surface into the gas phase, $E_{desorb} = (E_{surface} + E_{\alpha}^{gas\ phase}) - E_{surface+\alpha}$. For desorption, the energies are all based on the model described in Appendix \ref{app:energy_model}. The attempt frequencies, $\nu_\alpha$, are fixed for each species: $\nu_A = 6\times10^{12}/s$; $\nu_B = 4\times10^{12}/s$; $\nu_X = 6\times 10^{12}/s$; $\nu_{AX} = \nu_A\nu_X/(\nu_A + \nu_X) = 6\times10^{12}/s$; $\nu_{BX} = \nu_B\nu_X/(\nu_B + \nu_X) = 2.4\times10^{12}/s$; $\nu_{BX_2} = \nu_B\nu_X/(2\nu_B + \nu_X) =1.71\times10^{12}/s$. The formulas relating the attempt frequencies of the molecular species to the atomic attempt frequencies are not strongly suggested by theory, they were for simplicity taken to have a form inspired by geometric averages that leads to lower attempt frequencies for molecules because they are heavier than atoms. 

\item \emph{Diffusion of ions.}

Any ion is allowed to diffuse by hopping from the site it currently occupies to any unoccupied nearest-neighbor site, with rate $R_{hop}^{\alpha} = \nu_\alpha^{hop} e^{-E_{barrier}^\alpha /k_B T}$. We found it prohibitive to compute hopping barriers for enough adsorbate or defect states at the surface to fit our energy model. Therefore, $E_{barrier}^\alpha$ was taken from a simple formula that assumed the barrier shifted linearly from some average value based on difference in energy for the ion at the starting ($E_i^\alpha$) and ending ($E_j^\alpha$) sites, $i$ and $j$ respectively: $E_{barrier}^\alpha = \epsilon_0^\alpha + (E_j^\alpha - E_i^\alpha)/2$. The shift based on site energies ensures that the hop frequencies $i\rightarrow j$ and $j\rightarrow i$ obey detailed balance in a closed system \cite{PhysRev.37.405}.  The barrier shift, $\epsilon_0^\alpha$, for each species was chosen by taking the computed barrier for vacancy migration in the bulk and reducing it in an {\it ad hoc} manner (because barriers would likely be lower on a surface) to get some amount of diffusion for each species without diffusion becoming the dominant event: $\epsilon_A^0 = 0.27\ eV$, $\epsilon_B^0 = 0.54\ eV$, $\epsilon_X^0 = 0.5\ eV$.
Attempt frequencies, $\nu_\alpha$, were fixed for each species to be the same as for desorption attempts: $\nu_A = 6\times10^{12}/s$; $\nu_B = 4\times10^{12}/s$; $\nu_X = 6\times 10^{12}/s$.

\end{enumerate}

\subsection{Kinetic Monte Carlo algorithm for vapor deposited crystal growth}\label{kmc_processes}

This sub-section describes how crystal growth proceeds using the rates describe above using the rejection-free KMC time stepping algorithm \cite{gillespie1977exact} implemented in the SPPARKS kinetic Monte Carlo code \cite{webb_edmund_blackburn_iii_crossing_2009}.

\begin{enumerate}

\item[0.] (First step only): Initialize each lattice site in the simulation cell to be \texttt{occupied} (e.g., if there is a seed substrate) or \texttt{unoccupied}, with the species specified for occupied sites. Form a list of all events and associated rates for each lattice site. 

\item Choose three random numbers ($r_1$, $r_2$, $r_3$) from the uniform distribution on the unit interval. The first random number is used to select one of sites at which to perform an event, the second chooses one of the possible events at the chosen site, and the third gives a timestep $\tau = (1/a_0)\log{(1/r_3)}$, where $a_0$ is the sum of rates over all the events at all the sites in the system.

\item Change the lattice occupancies at any sites that are affected by the selected event, and advance the time by $\tau$.

\item Update the list of possible events for the site chosen in step 1, and any of its affected neighbors, next-nearest neighbors, etc., because changes to a site might affect what events are possible at nearby sites.

\item If the time is less than the maximum time chosen for the \gls{kmc} simulation, return to step 1.

\end{enumerate}

\section{Local-neighborhood energy model}\label{app:energy_model}

\subsection{\gls{dft} calculations of defect and adsorbate energetics}\label{dft_site_energies}

We used plane-wave \gls{dft} to compute the energies of ion vacancies and ion adsorbates for orthorhombic $CsPbI_3$, stable at ~300 K, using the X-ray diffraction structure for a temperature of 325 K published by Sutton et al. \cite{sutton2018cubic}. All energy calculations were performed with plane-wave \gls{dft} using the Vienna Ab Initio Software Package (VASP), version 5.4.4 \cite{kresse1993ab,kresse1994ab,kresse1996efficiency,kresse1996efficient,blochl1994projector,kresse1999ultrasoft}. Electronic structure calculations of system energies were computed using the PBE form of generalized gradient exchange-correlation functional,\cite{PhysRevLett.77.3865} supplemented by the Grimme's D3 van der Waals correction with Beck-Johnson damping \cite{jcc.21759} and default damping parameters (VASP setting IVDW=12). The convergence parameters and k-point meshes were generated using the Pymatgen \cite{Ong2013} MPRelaxSet command with the default reciprocal k-point density of 64.

The defect energies of individual neutral $Cs$, $Pb$, and $I$ vacancies in bulk $CsPbI_3$ and from $CsPbI_3$ surfaces were calculated as the difference between the energy of the lattice with vacancy and the total energy of the lattice (bulk or surface) system without defects. We write this schematically as, \begin{eqnarray}
E_{defect} = E(CsPbI_3\ \mathrm{lattice\ with\ vacancy}) - E(CsPbI_3\ \mathrm{lattice}) ,
\label{e_vacancy}
\end{eqnarray} where $E(\mathrm{structure})$ is the total computed energy of the structure indicated in parentheses. Adsorption energies of individual $Cs$, $Pb$, and $I$ atomic adsorbates on surfaces were calculated as the difference between the energy of the surface with an adsorbate and the energy of a perfect (110) surface,\begin{eqnarray}
E_{adsorbate} = E(CsPbI_3\ \mathrm{(100)\ slab\ with\ adsorbate}) - E(CsPbI_3\ \mathrm{(100)\ slab}) 
\label{e_adsorbate}
\end{eqnarray}
Note that the defect and adsorbate energies, which are what we fit in our lattice energy model \ref{enmodel}, are not the energies to move an atom to or from the gas phase, respectively, because the desorption and adsorption energies must also subtract off the isolated-atom energy. This is done as a convenience to avoid adding a constant energy independent of $\{n_{Cs},n_{Pb},n_I\}$ to each of the $F_\alpha$. The gas-phase energy is explicitly subtracted from the defect energy in our KMC implementation of processes 2 in Appendix \ref{kmc_processes}.

For the adsorption and desorption from the orthorhombic surface, we formed a 1x1x8 supercell slab of the (110) surface, with the bottom six layers constrained to the experimental positions and relaxation allowed for the top two layers and any adsorbate. The energy of the neutral slab was computed and we computed adsorbates and defects on or in two possible surface layers, one with $CsI$ stoichiometry and the other with $PbI_2$ stoichiometry. The $CsI$ surface had a total of 4 unique $Cs$ and 4 unique $I$ sites and the $PbI$ surface had a total of 4 unique $Pb$ sites and 8 unique $I$ sites, for a total of 20 surface defect energies computed (4 $Cs$, 4 $Pb$, and 12 $I$). We also identified 12 adsorption locations for the $CsI$ surface and 8 for the $PbI$ surface, for a total of 60 distinct adsorption energies (20 for $I$, 20 for $Pb$, and 20 for $I$). In addition to the 20 defect and 60 adsorption energies described above, we also included defect energies from the bulk orthorhombic lattice, 2 $Cs$, 2 $Pb$, and 2 $I$ defects, for a total of 86 energies used to fit the energy model described below. All the energies are available in Tables S1-S6 of the Supplementary Information.

{\it A note on corrections for charges and dipoles:} For all calculations, the adsorbates and defects were taken to be charged (Cs$^+$, Pb${^2+}$, I$^-$). With charged defects or adsorbates, corrections to the energy due to the net charge and the surface dipole give the most accurate results. \cite{PhysRevB.46.16067,PhysRevB.51.4014} It {\it is} possible to compare relative energies of similar charged systems without the corrections (e.g., two different Cs$^+$ surface defect sites) because the charge and dipole corrections will be similar to each other, but energies among adsorbates or defects with different charges (e.g., Pb$^{2+}$ vs I$^-$) will not be reliably compared because dipole and monopole corrections will necessarily be similar to those with other ions. We found, however, that such corrections were computationally very expensive (more so because the orthorhombic lattice means simplifications for a square cross section cannot be used). Hence, instead of computing charge and dipole corrections, we used a large vacuum region above and below the slab to reduce the magnitude of the error. In test cases, the monopole and dipole corrections were of order 1 eV, which is similar to the horizontal spread seen in the separated clusters of energies shown in Fig.~\ref{fig:en_model}, so we considered this approximation acceptable for the semi-quantitative purposes of this paper.

\subsection{Neighbor-density energy model}\label{enmodel}

Using the bulk and surface defect energies and the surface adsorption energies, we constructed a simplified model for the energy of each ion in the $CsPbI_3$ lattice based on how many of each species occupy neighboring lattice sites. In this model, the energy of an atom of type $\alpha \in (Cs,Pb,I)$ at a site $i$ is defined to be a function of the number of occupied neighbors of each type, $E_i^\alpha = F_\alpha(\{n_{Cs}, n_{Pb}, n_{I}\})$, where $n_\beta$ is the number of ions of type $\beta$ within a cutoff distance $R_{cut}$ of site $i$. We tested two values of $R_{cut}$, 5.0 \AA, which includes nearest neighbors of each site, and 5.9 \AA, which includes some next nearest neighbors, allowing for, e.g., $Pb-Cs$ interactions and not just $Pb-I$ and $Cs-I$ nearest-neighbor interactions in the perfect perovskite lattice. We obtained much better fits with $R_{cut}$ = 5.9 \AA\ so this cutoff was used for the final model. We fitted both  linear and linear+quadratic forms of $F$ and found that the best fit was obtained from three linear+quadratic functions, one for each $\alpha\in\{Cs,Pb,I\}$,
\begin{equation}
F_\alpha(n_{Cs}, n_{Pb}, n_{I}) = \sum_{\beta\in\{Cs,Pb,I\}} n_{\beta} A_{\alpha;\beta} + \sum_{\beta,\gamma\in\{Cs,Pb,I\}} n_\beta n_\gamma B_{\alpha;\beta\gamma}\ .
\label{quadfit}\end{equation}
Note that this form for the $F_\alpha$ says that an ion at site $i$ has an energy that depends only the neighbor counts, it carries no information about inter-ion distances or whether the neighbors occupy particular lattice sites. This means it does not privilege occupation of the ``correct'' locations by each ion type.
The three functions are coupled because we symmetrized the interaction terms, $A$ and $B$, such that $A_{\alpha;\beta} = A_{\beta;\alpha}$ and $B_{\alpha; \beta\gamma} = B_{\alpha; \gamma\beta}$, which gave a total of 24 parameters to jointly fit 86 energies across the three functions, $F_{Cs}$, $F_{Pb}$, $F_{I}$. The fit was performed using the default trust region reflective algorithm of the \texttt{scipy.optimize.least\_squares} routine in  SciPy version 1.14.1 \cite{2020SciPy-NMeth}. The modeled energies are plotted versus computed energies in Fig.~\ref{fig:en_model}. The fitted parameters, $A$ and $B$, are given in Table S7 of the Supplementary Information.

\begin{figure}[!htbp]
    \centering
    \includegraphics[width=0.8\linewidth]{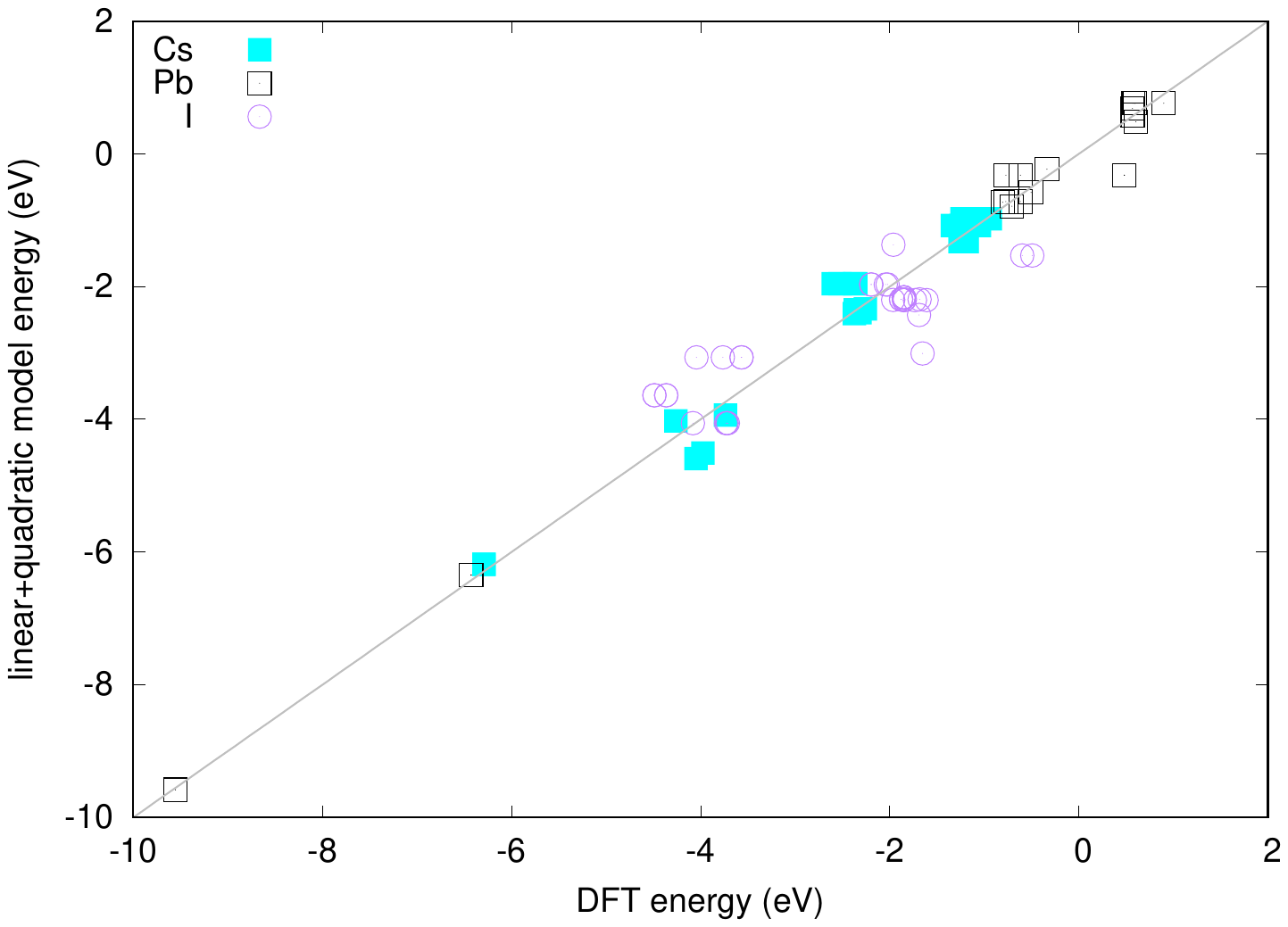}
    \caption{Modeled defect and adsorbate site energies from Eq.~\ref{quadfit} versus computed energies (Appendix \ref{dft_site_energies}) for $Cs$ (filled cyan squares), $Pb$ (black squares), and $I$ (purple circles). The gray line shows the ideal $E_{model} = DFT\ energy$ result to guide the eye.}
    \label{fig:en_model}
\end{figure}

\section{Comparison with a constant-flux model}\label{app:constant_flux}

The directly coupled model in Figure~\ref{fig:accuracy} exhibits a simple evolution: an initial high-deposition phase rapidly depletes partial pressures, followed by an equilibrated phase with nearly constant deposition rates and partial pressures. This motivates testing whether such behavior can be described by a model with constant flux of each species out of the domain and onto the substrate. The benefit of a constant-flux model is that the coupling between the macro and microstates becomes a constant boundary condition in a purely fluid dynamical model, and the expensive KMC simulations need not be called.

\begin{figure}
    \centering
    \includegraphics[width=1.0\linewidth]{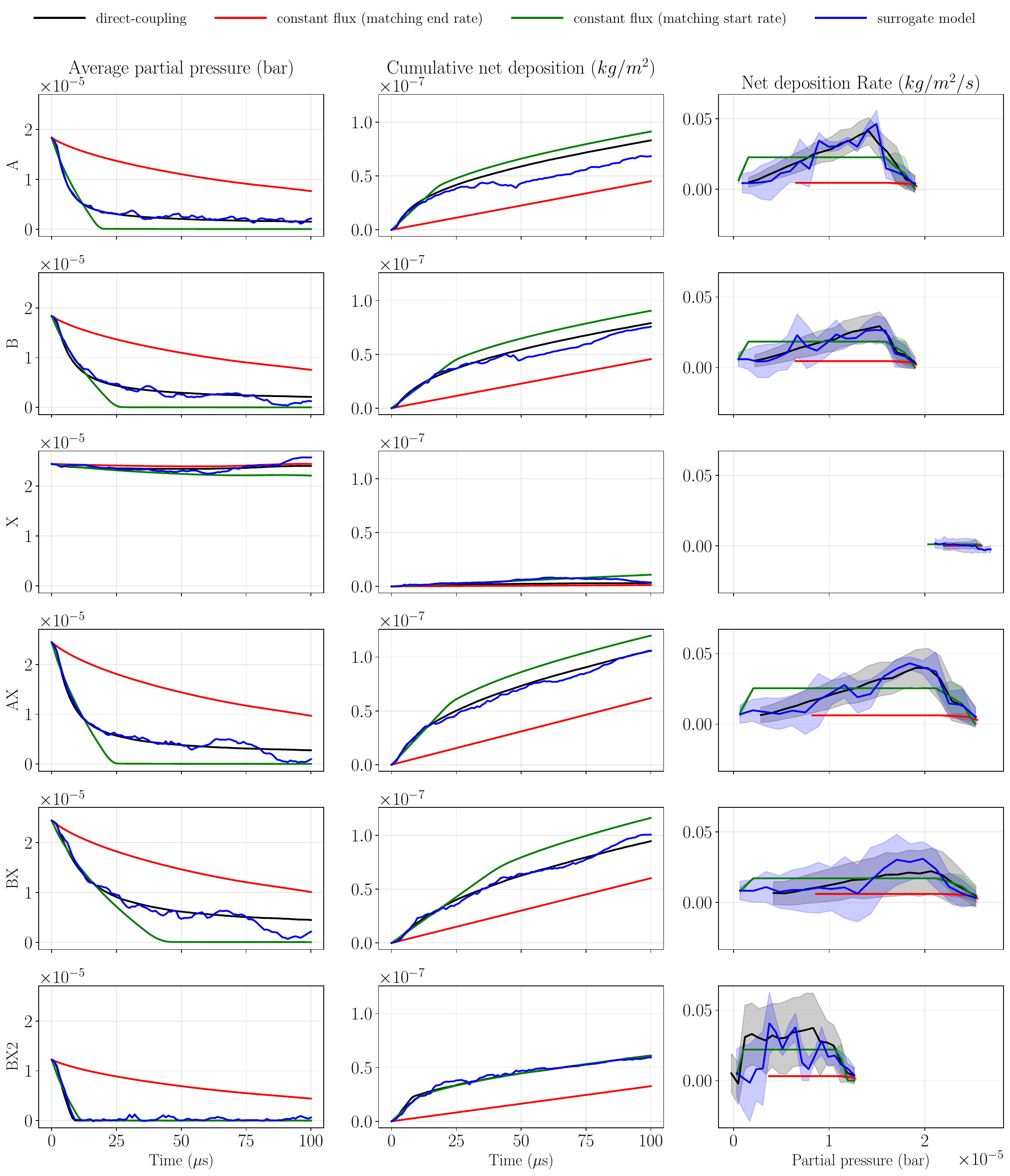}
    \caption{The directly-coupled model and surrogate model at the lowest variance threshold (0.02) (see also Figure~\ref{fig:accuracy}) are compared against pure fluid-dynamical models with a constant prescribed flux for each species at the substrate boundary. The green curve matches the starting rate (averaged over the first 5~$\mu$s) of the directly coupled model, and the red curve matches the end rate (averaged over the last 25~$\mu$s) of the directly-coupled model.}
    \label{fig:constflux}
\end{figure}

In Figure~\ref{fig:constflux}, we compare constant-flux models against the directly-coupled model and the best performing surrogate model (with the lowest variance threshold, 0.02). Because the appropriate flux is not known a priori, we consider a high-flux case matched to the diagnosed deposition rates of the direct-coupled simulation's initial fast phase (first 5~$\mu$s) and a low-flux case matched to the equilibrated phase (last 25~$\mu$s).

The low-flux model matches the equilibrium deposition rate of the directly-coupled simulation but misses the initial rapid deposition and thus consistently underestimates cumulative net deposition. The high-flux model instead depletes each boundary cell until its partial pressure reaches zero, after which deposition is limited by transport from the interior. It therefore generally overestimates cumulative deposition, except for BX2, which is also nearly fully depleted in the directly coupled model. For the other species, from the directly-coupled simulation, we understand that as the partial pressure of that species reduces, so too does the deposition rate; hence, each species is not depleted fully from the boundary grid cell in the equilibrium phase, and cumulative net deposition is lower than the high-flux model prediction. By construction, a constant-flux model will never account for this process.

Constant flux may therefore be adequate when a species is known to remain nearly depleted at the boundary and deposition is transport-limited. Otherwise, choosing an accurate flux requires information from a reliable coupled simulation. A simplified and deterministic model for species flux may still be a useful replacement for microsimulations. The right column of Figure~\ref{fig:constflux} suggests approximating deposition rates as a piecewise linear function of partial pressure, but setting this approximation still requires initial data from a reliable coupled simulation.

\end{appendices}

\newpage
\bibliography{refs}

\end{document}